\documentclass[twocolumn]{aastex62}
\pdfoutput=1 
\usepackage{amstext,amsmath}
\usepackage[T1]{fontenc}
\usepackage[figure,figure*]{hypcap}
\usepackage{comment}

\usepackage{amsmath}
\usepackage[bb=boondox]{mathalfa}
\usepackage{graphicx}
\graphicspath{ {./} }

\shorttitle{Modelling the Galactic Foreground and Beam Chromaticity}
\shortauthors{Hibbard et al.}

\begin{document}

\title{Modelling the Galactic Foreground and Beam Chromaticity for Global 21-cm Cosmology}

\correspondingauthor{Joshua J. Hibbard}
\author{Joshua J. Hibbard}
\affiliation{Center for Astrophysics and Space Astronomy, Department of Astrophysical and Planetary Science, University of Colorado Boulder, CO 80309, USA}

\author{Keith Tauscher}
\affiliation{Center for Astrophysics and Space Astronomy, Department of Astrophysical and Planetary Science, University of Colorado Boulder, CO 80309, USA}
\affiliation{Department of Physics, University of Colorado, Boulder, CO 80309, USA}

\author{David Rapetti}
\affiliation{NASA Ames Research Center, Moffett Field, CA 94035, USA}
\affiliation{Research Institute for Advanced Computer Science, Universities Space Research Association, Mountain View, CA 94043, USA}
\affiliation{Center for Astrophysics and Space Astronomy, Department of Astrophysical and Planetary Science, University of Colorado Boulder, CO 80309, USA}

\author{Jack~O.~Burns}
\affiliation{Center for Astrophysics and Space Astronomy, Department of Astrophysical and Planetary Science, University of Colorado Boulder, CO 80309, USA}

\email{Joshua.Hibbard@colorado.edu}

\begin{abstract}
In order to characterize and model the beam-weighted foreground for global 21-cm signal experiments, we present a methodology for generating basis eigenvectors that combines analytical and observational models of both the galactic spectral index and sky brightness temperature with simulations of beams having various angular and spectral dependencies and pointings. Each combination creates a unique beam-weighted foreground. By generating eigenvectors to fit each foreground model using Singular Value Decomposition (SVD), we examine the effects of varying the components of the beam-weighted foreground. We find that the eigenvectors for modelling an achromatic, isotropic beam---the ideal case---are nearly identical regardless of the unweighted foreground model used, and are practicably indistinguishable from polynomial-based models. When anisotropic, chromatic beams weight the foreground, however, a coupling is introduced between the spatial and spectral structure of the foreground which distorts the eigenvectors away from the polynomial models and induces a dependence of the basis upon the exact features of the beam (chromaticity, pattern, pointing) and foreground (spectral index, sky brightness temperature map). We find that the beam has a greater impact upon the eigenvectors than foreground models. Any model which does not account for its distortion may produce RMS uncertainties on the order of $\sim 10 - 10^3$ Kelvin for six-parameter, single spectrum fits. If the beam is incorporated directly using SVD and training sets, however, the resultant eigenvectors yield milli-Kelvin level uncertainties. Given a sufficiently detailed description of the sky, our methodology can be applied to any particular experiment with a suitably characterized beam for the purpose of generating accurate beam-weighted foreground models.
\end{abstract}


\section{Introduction}
\label{sec:introduction}

Of all challenges arrayed against studies of primordial neutral hydrogen cosmology, the problem of modelling the diffuse galactic foreground remains one of the most singular. It is bright, dynamic, and ubiquitous. In order to study the elusive sky-averaged (global) 21-cm cosmological signal and hence the thermal evolution of the Universe from the Dark Ages (DA) to the Epoch of Reionization (EoR), one must be able to separate this contaminating galactic foreground from the signal, which is itself four to six orders of magnitude fainter (\cite{Furlanetto:06}). This precision cosmology seemingly requires accurate sky brightness temperature maps of the low-radio frequency sky, of which, in the frequency ranges of interest, there is currently a considerable dearth. Below 1 GHz, there are merely two temperature maps with greater than $\sim 96 \% $ sky coverage, that of \cite{Haslam:82} and \cite{Guzman:2011}. Consequently, numerous experiments in DA and EoR astrophysics rely upon extrapolating these sky brightness temperature maps to model the foreground, yet the errors in these maps are on the order of ten percent, far from ideal for precision cosmology. Nonetheless, in the absence of temperature maps with markedly smaller (or even quantified) errors, the burden of rigor must fall upon the method of modelling and subsequent analysis for signal extraction. Great strides have been made in the effort to generate full-sky, low-frequency foreground temperature maps, as well as to combine them using a Principal Component Analysis to build the Global Sky Model \cite[GSM;][]{Zheng:2017}. The latter will be used extensively throughout this work.

There are currently several experiments and mission concepts to detect the global 21-cm signal, all of which must grapple with the problem of the bright foreground: the Experiment to Detect the EoR Signature \citep[EDGES;][]{Bowman:18,Monsalve:17,Monsalve:2019}, the Shaped Antenna measurement of the background RAdio Spectrum \citep[SARAS;][]{Patra:2013,Singh:2017}, the Sonda Cosmol\'ogica de las Islas para la Detecci\'on de Hidr\'ogeno Neutro \citep[SCI-HI;][]{Voytek:2014}, the Zero-spacing Interferometer Measurements of the Background Radio Spectrum \citep[ZEBRA;][]{Mahesh:2014}, the Large-aperture Experiment to detect the Dark Ages \citep[LEDA;][]{Bernardi:2015,Bernardi:2016,Price:2018}), the Broadband Instrument for Global HydrOgen ReioNization Signal \citep[BIGHORNS;][]{Sokolowski:2015}, and the Dark Ages Polarimeter PathfindER \citep[DAPPER;][]{Burns:2020b,Burns:2020a}.

Modelling the galactic foreground for global 21-cm experiments has typically consisted of extrapolating temperature maps to lower frequencies and correcting them for antenna beam systematics \citep[applying for example the beam chromaticity factor used in][]{Bowman:18}, but neglecting the finer spatial and spectral structures in the foreground. Instead, the overall, relatively smooth spectral form of the foreground is assumed to be well characterized by polynomial-based models, such as a power-law times a polynomial or logarithmic polynomial {\cite[see e.g.][]{SathyanarayanaRao:17, Monsalve:17,Bowman:18}}. However, for global 21-cm measurements, the foreground is not a single power law, but dominated instead by many power laws averaged together.

Moreover, the beams used in most global 21-cm experiments display prodigious dependencies of the beam's angular structure upon frequency. Consequently, at each frequency the simple sum of power laws changes due to differential weighting of each point in the sky across the bandwidth. With a perfectly achromatic beam, each point in the foreground would contribute to a weighted, averaged antenna response (including foreground and signal), with the weights contingent only upon the angular dependence of the beam and the pointing of the antenna. In the case of a chromatic beam, however, the weights of each angle in the sky are frequency dependent, making the full foreground spectrum more complicated than a simple weighted average of the spectra of the foreground in each direction. Thus, beam chromaticity tends to confuse the spatial and the spectral, as different weighted skies or sums of power-laws will be seen at each frequency. Indeed, unless correspondingly accurate simulations or measurements of receiver beams are generated, the precision of any detection lies, once more, wholly with the method of modelling. For without properly accounting for both foreground contamination and beam distortions to high degrees of accuracy, the beam-weighted foreground model may contain spurious residuals relative to the true beam-weighted foreground, see \cite{Tauscher:2020a} (hereafter T20a).

We therefore confront the following questions: what is the spectral and spatial structure of the foreground? How does the beam and its chromaticity interact with the spatial foreground dependence to distort the spectral structure of the foreground and subsequently affect the beam-weighted models? What features of the beam-weighted foreground are the most important to account for in a model?

Independently of the overlap with the global 21-cm signal model, having a model representative of the beam-weighted foreground variations for a particular experiment with a specific beam is key for our pipeline. Because the 21-cm signal and foreground are not expected to be orthogonal, they must be fit simultaneously. If instead the foreground is fit and subtracted before fitting the signal, the foreground fit will remove signal power and the signal fit will be biased. Our pipeline, as laid out in \cite{Tauscher:18}, \cite{Rapetti:2019}, and \cite{Tauscher:20} creates basis vectors that are optimal for fitting signal and foreground training sets. The simultaneous fitting procedure of the pipeline produces uncertainties that properly account for the covariance between signal and foreground parameters.\footnote{The overlap between signal and foreground models, as determined by their training sets, can be decreased by using polarization and time-dependence \citep[see Paper III of our pipeline,][]{Tauscher:2020a}}.

Thus, in this paper we focus on a methodology for creating beam-weighted foreground models via realistic training sets and on comparing the underlying characteristics used to build them based on the impact that these have on the eigenmodes forming such models.

In such a pursuit, we present a technique for analyzing the beam-weighted foreground using pattern recognition by applying Singular Value Decomposition (SVD) on training sets, building upon the work of \cite{Tauscher:18}, hereby referred to as T18. The procedure of T18 starts by generating a simulated set of all possible ways in which a particular systematic (or the signal) can vary by simulating a training set of that systematic. In our case, the systematic effect of interest is the beam-weighted foreground. Next, SVD is used to generate singular values (eigenvalues or importances) and singular vectors (or eigenmodes). If the training set accurately represents all of the possible ways in which the beam-weighted foreground can vary, then its SVD eigenmodes form an optimal basis with which to fit the curves in the set, with the eigenmodes ordered by importance.

In Section \ref{Methods} we build models of the foreground by combining analytical and observational models of the spectral index with sky brightness temperature maps. Each unique realization of the two variables is then used to create a distribution of possible foregrounds, which are then drawn from to assemble a training set describing a possible galactic foreground. Section \ref{Results_1} treats the ideal case of the monopole-beam, or a perfectly achromatic, isotropic beam and its resultant eigenmodes, which we find are similar to the polynomial-based models mentioned above. In Section~\ref{Results_2} we apply examples of more realistic, chromatic beams to duly distort the foreground, and in Section~\ref{Discussion} we perform a quantitative analysis of the effects of these beams' distortions on the SVD eigenmodes of the beam-weighted foreground with respect to those of the optimal basis. In this section, we also calculate the level of milli-Kelvin residuals to be expected when using a given SVD basis to describe a certain training set, and examine the importance---in terms of root-mean-square (RMS) residuals---and the relative effect of varying the 
beam and foreground characteristics upon the beam-weighted foreground model and eigenmodes. We conclude in Section~\ref{sec:conclusions}.

\section{Methods: \\ Galaxy Models}
\label{Methods}

In the low frequency ranges of interest to DA and EoR cosmology, synchrotron radiation is the brightest source of contamination (\cite{Furlanetto:06}). The latter is well-modelled by a power law, (assuming the underlying electron energy distribution generating the radiation also follows a power-law, see \cite{Condon:16}):
\begin{equation}
    T_{sky}(\nu,\theta,\phi) = T_{map}(\theta,\phi) \Bigg( \frac{\nu}{\nu_o} \Bigg)^{\beta(\theta,\phi,\nu)},
    \label{powerlaweq}
\end{equation}
where $\beta$ is the relatively well-measured galactic spectral index (GSI), which can vary spatially and spectrally\footnote{Frequency dependence of the spectral index causes what is commonly referred to as spectral curvature.}, $\nu_o$ is a reference frequency, and $T_{map}$ gives the sky brightness temperatures for the galaxy as a function of galactic coordinates at the reference frequency. There are thus two primary ways to change the sky model, $T_{sky}$: vary the temperature maps ($T_{map}$), or vary the spectral index ($\beta$). For this work, we shall consider $\beta$ to vary only spatially.

However, the temperature actually measured by the antenna weights each spatial coordinate according to the pattern and chromaticity of the beam:
\begin{equation}
    T_{ant}(\nu) = \frac{\int_{4\pi} T_{sky}(\nu,\theta,\phi)*B(\nu,\theta,\phi) d\Omega}{\int_{4\pi} B(\nu, \theta,\phi) d\Omega}. 
    \label{eq:beam-weighted-foreground-measurement-equation}
\end{equation}
Here, $B$ is the beam's chromaticity, a function that weights each spatial coordinate $d\Omega$. Typically the beam is normalized so that $\int B(\nu,\theta,\phi)\ d\Omega=1$, although it still changes with frequency $\nu$. A fully-realized model must account for the effect of the beam.

The discrete version of equation \ref{eq:beam-weighted-foreground-measurement-equation}, useful for modelling and mapping in HEALPIX (\cite{Gorski:2005}), is
\begin{equation}
    T_{ant}(\nu) = \frac{\sum_{i=1}^{N_{pix}} T_{sky}(\nu,i)*B(\nu,i)}{\sum_{i=1}^{N_{pix}}B(\nu,i)},
    \label{discreteskytemp}
\end{equation}
where $i$ denotes each pixel, and $N_{pix}$ gives the total number of pixels measured by the beam.

In the next section we generate analytical maps of the galactic foreground ($T_{sky}$ in Equation \ref{discreteskytemp}) by modelling the two ways in which it may vary: that is, spectrally through changes in $\beta(\theta,\phi)$ and spatially through changes in $T_{map}(\theta,\phi)$, the temperature map used for the sky.

Then, in Section \ref{interpolated-observational-maps} we compare these analytical maps to observational ones derived from published temperature maps (such as the GSM).

In later sections, we use these maps and Equation \ref{discreteskytemp} to simulate the beam-weighted foregrounds.

\subsection{Analytical Spectral Index Models}
\label{sec-analytical-spectral-index-models}

The work of \cite{Guzman:2011}, \cite{Mozdzen:2019}, \cite{Mozdzen:2017}, \cite{Rogers:2008}, among others, have shown that the most significant spatial variations in the GSI at these low radio frequencies are attributable to contributions from observations with the galactic plane overhead. In general, the appearance of the galactic plane causes a flattening of the spectral index. As such, the analytical GSI maps we utilize assume that the most significant changes in the GSI are captured by co-latitude variation in galactic coordinates.

The first model assumes that the GSI variation can be modelled as a Gaussian function of co-latitude ($\theta$) centered at the galactic plane and falling off 10 degrees above and below the center (see Equation \ref{gaussmodel}). The second and third models assume that the GSI varies with galactic co-latitude as $\sin^2(\theta)$, with the second assuming a Gaussian distribution in the offset and magnitude of variation parameters $O$ and $M$, and the third assuming constant magnitudes and variation, but Gaussian noise (constant perturbations) added to the entire sky (see Equations~\ref{sin2model} and~\ref{sin2pmodel}, respectively). We hereafter refer to each of these analytical models as Gaussian, Sine-squared, and Perturbed, respectively. The governing equations for these models are the following:

\begin{subequations}
\begin{align}
    \beta_{G}(\theta) &= O + M  \exp \Bigg[ \frac{-(\theta - \pi/2)^{2}}{2 \sigma^2} \Bigg], \label{gaussmodel} \\
    \beta_{S}(\theta) &= O + M \sin^2{\theta}, \label{sin2model} \\
    \beta_{P}(\theta) &= O_{o} + M_{o} \sin^2{\theta} + n. \label{sin2pmodel}
\end{align}
\end{subequations}
Here, $\theta$ is the galactic co-latitude coordinate, and $\sigma$ sets the scale of the Gaussian model width, which we take to be about 10 degrees. 

Assuming, for instance, that the GSI flattens out to -2.1 near the galactic center and attains a maximum value of -2.5 near the poles, then $O = -2.5$ and $M = 0.4$. To create a wide set of GSI maps given any particular offset and magnitude, we approximate both $O$ and $M$ as coming from a normal distribution with a standard deviation of $0.01$ for the Gaussian and Sine-Squared models. The Perturbed model assumes that the offset and magnitude are constant. Instead, a single value of Gaussian noise $n\sim N(0,0.01)$ is added to every coordinate, where $N(\mu,\sigma)$ refers to a normal distribution with mean $\mu$ and standard deviation $\sigma$, shifting the entire GSI co-latitude curve up or down equally in the spectral index amplitude.

\begin{figure}
    \centering
    \includegraphics[width=0.46\textwidth]{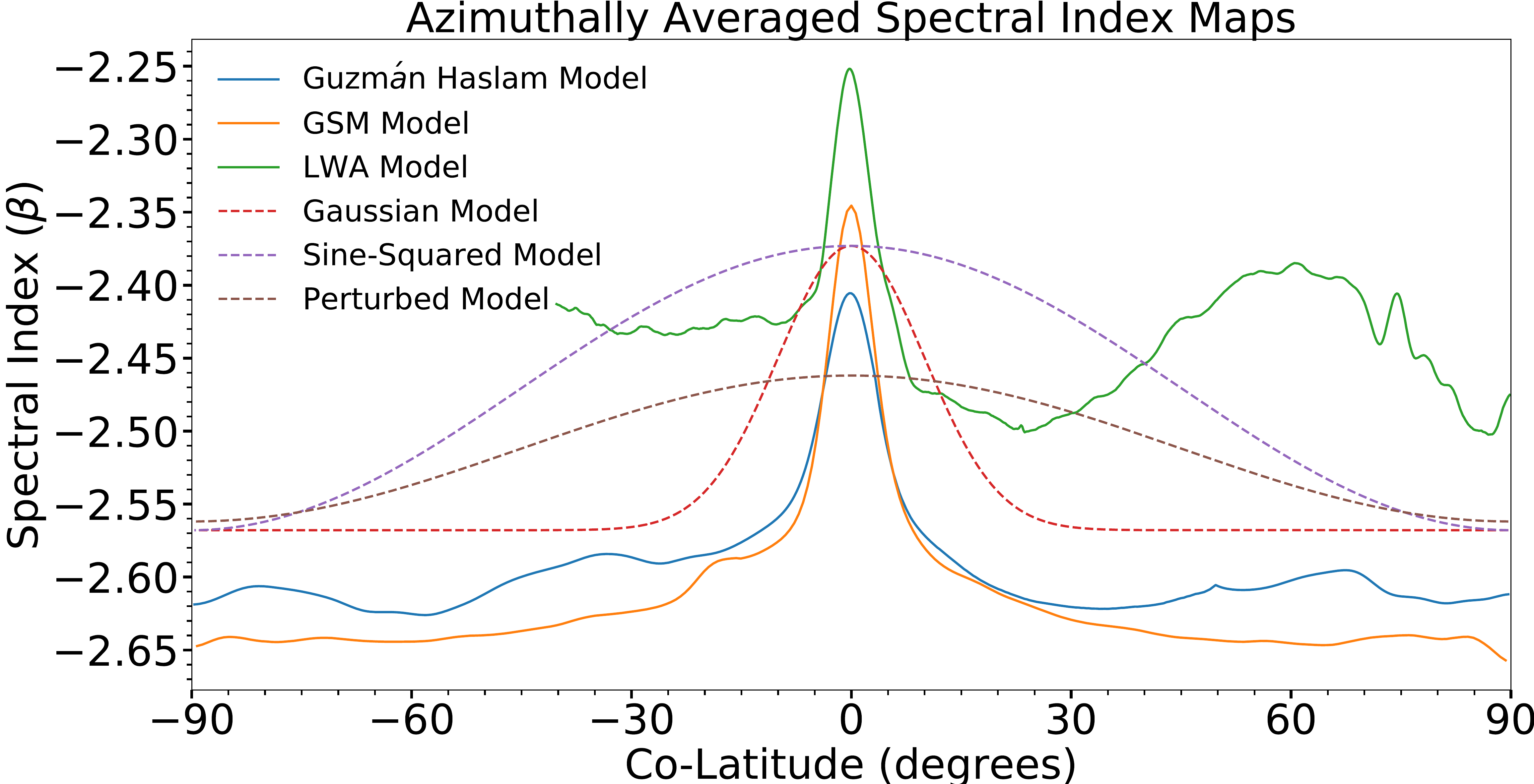}
    \caption{Example spectral index curves (encompassing the frequency range 40-120 MHz), averaged over longitude to show the co-latitude dependence of each model. Solid lines represent Interpolated Observational models, and dashed lines represent analytical models. The most significant variation in the spectral index occurs at the galactic plane $\theta = 0$, motivating the functional forms of the analytical spectral index models of Equations \ref{gaussmodel} - \ref{sin2pmodel}. Note that the observational maps contain contributions from bright point-sources, while the analytical ones do not.}
    \label{fig:example_si}
\end{figure}

For this work, we used two sets of offset and magnitude parameters for each of the three analytical models above to generate GSI maps using the spectral index curves shown as dashed lines in Figure \ref{fig:example_si}. The first set uses $O=-2.5$ and $M=0.4$ for the distribution means, while the second set uses the spectral index range reported by \cite{Mozdzen:2019}, for which approximately $O=-2.59$ and $M=0.13$. This is useful for comparison purposes and to show the effect of GSI plane-to-pole spectral index range upon foreground modelling.

\subsection{Interpolated Observational Spectral Index Maps}
\label{interpolated-observational-maps}

By fitting a power law of the form of Equation \ref{powerlaweq} between two or more published sky brightness temperature maps, it is possible to create interpolated observational (IO) GSI maps. Such a fit is computed at every pixel, assuming the maps have (or have been converted to have) the same number of pixels, giving galactic sky maps of the spectral index (instead of temperature).

The following published sky brightness temperature maps were used to create three IO GSI maps by interpolating between the maps listed in each bullet-point:

\begin{itemize}
    \item The 45 MHz map of \cite{Guzman:2011} and the 408 MHz map of \cite{Haslam:82}
    \item Global Sky Model (GSM) maps generated between 40 and 120 MHz with 5 MHz spacing from \cite{Zheng:2017}
    \item Nine Long Wavelength Array (LWA1) low frequency sky survey maps between 35 and 80 MHz from \cite{Dowell:2017}.
\end{itemize}

\begin{figure}
    \centering
    \includegraphics[width=0.46\textwidth]{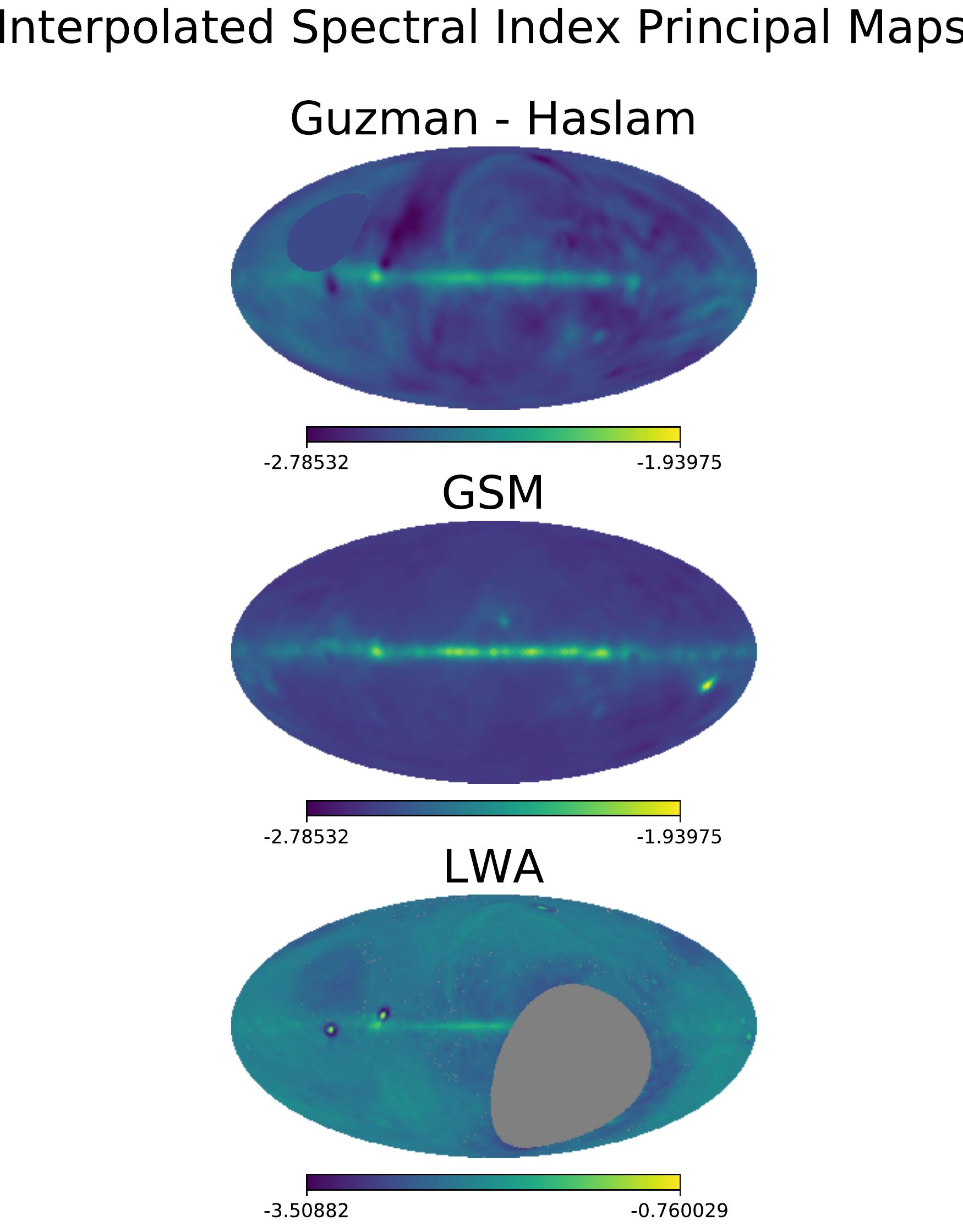}
    \caption{Principal interpolated GSI maps. The colored values show the interpolated spectral index at various places in the galactic sky. The hole in the Guzman-Haslam is filled with the average surrounding value, while the southern hemisphere that is missing from the LWA map is removed before further calculations are made.}
    \label{fig:io_master_maps}
\end{figure}

We denote the resulting IO GSI maps as ``principal'' and represent them with the symbol $\beta_{IO}$. Mollweide projections of these three principal maps are shown in Figure \ref{fig:io_master_maps} and their longitudinally averaged values are shown by the solid lines in Figure~\ref{fig:example_si}. We note also that the GSM incorporates both the Haslam and Guzm\'an maps, in addition to contributions from several other partial low-frequency maps, filling in some gaps in spatial coverage.

In \cite{Guzman:2011}, they derive corrections to their extrapolated temperature maps in the form of zero-level and extragalactic temperature corrections by a large literature survey and subsequent fit. Their temperature maps are then recalculated using these corrections.  As we are interested in all of the possible ways in which these IO maps may vary across the galactic sky (for the purpose of eventually building training sets for the SVD analysis), we may approximate these derived corrections as corresponding to values for the errors in the IO Guzm\'an-Haslam and GSM maps. Using the maximum error that \cite{Guzman:2011} report between their purely uncorrected, extrapolated GSI map, and purely corrected, extrapolated GSI map, which is $\beta \pm 0.03$, we build two sets of IO GSI maps. To each of these maps, we also apply the following transformation:

\begin{equation}
    \beta(\theta,\phi) = \beta_{IO} (\theta,\phi) + n(\theta,\phi),
\end{equation}
where $n(\theta,\phi) \sim N(0,0.01)$ is a normal distribution and represents an independent perturbation at each coordinate (or pixel) and is different for each map in the set. Using either the GSM or Guzm\'an-Haslam IO principal map, we run this process many times to produce ``perturbed'' spectral index maps, each grouped into a set according to the principal map being perturbed. To construct a proper GSM training set we would draw curves using the reported parameter covariances, but since this information has not yet been published, here we employ the method described above to generate the GSM training set used in this work.

As the LWA1 Low-Frequency Sky Survey reports errors on their various temperature maps, we can use these to construct covariance matrices that describe the temperature standard deviation at each pixel. With these in hand, we then construct an LWA set of maps according to the expression

\begin{equation}
    \beta_{LWA}(\theta,\phi) = \beta_{IO,LWA}(\theta, \phi) + n \sigma(\theta,\phi),
\end{equation}
where $\beta_{IO,LWA}$ represents the principal IO LWA map, derived by fitting the 9 temperature maps to a power law (as with the GSM and Guzm\'an-Haslam before), $n$ is taken from a unit normal distribution, and $\sigma(\theta,\phi)$ is the derived spectral index covariance from the reported map errors and the power law fit. The places without reported temperatures in the LWA1 survey (pixels south of 40 degrees latitude in celestial coordinates) are removed from each map before any further analysis.

\subsection{Constant Spectral Index Maps} \label{sec:constant-spectral-index-maps}

In addition, we generated GSI maps with a constant spectral index for comparison. The following published spectral index ranges were used to generate the spectral index ranges used in creating the Constant GSI maps:

\begin{itemize}
    \item $-3.5$ to $-1.5$: \cite{Eastwood:2018}
    \item $-2.59$ to $-2.22$: \cite{Dowell:2017}
    \item $-2.7$ to $-2.1$: \cite{Guzman:2011}
    \item $-2.59$ to $-2.46$: \cite{Mozdzen:2019}.
\end{itemize}

For our analysis purposes, spectral indices were drawn randomly (using a uniform distribution) from the ranges of these four sets and were then combined into a single Constant GSI map set. To be clear, each galactic foreground generated for the Constant GSI set has a constant spectral index across the entire galactic sky.

\subsection{Sky Temperature Maps}

\begin{figure*}
    \centering
    \includegraphics[width=0.96\textwidth]{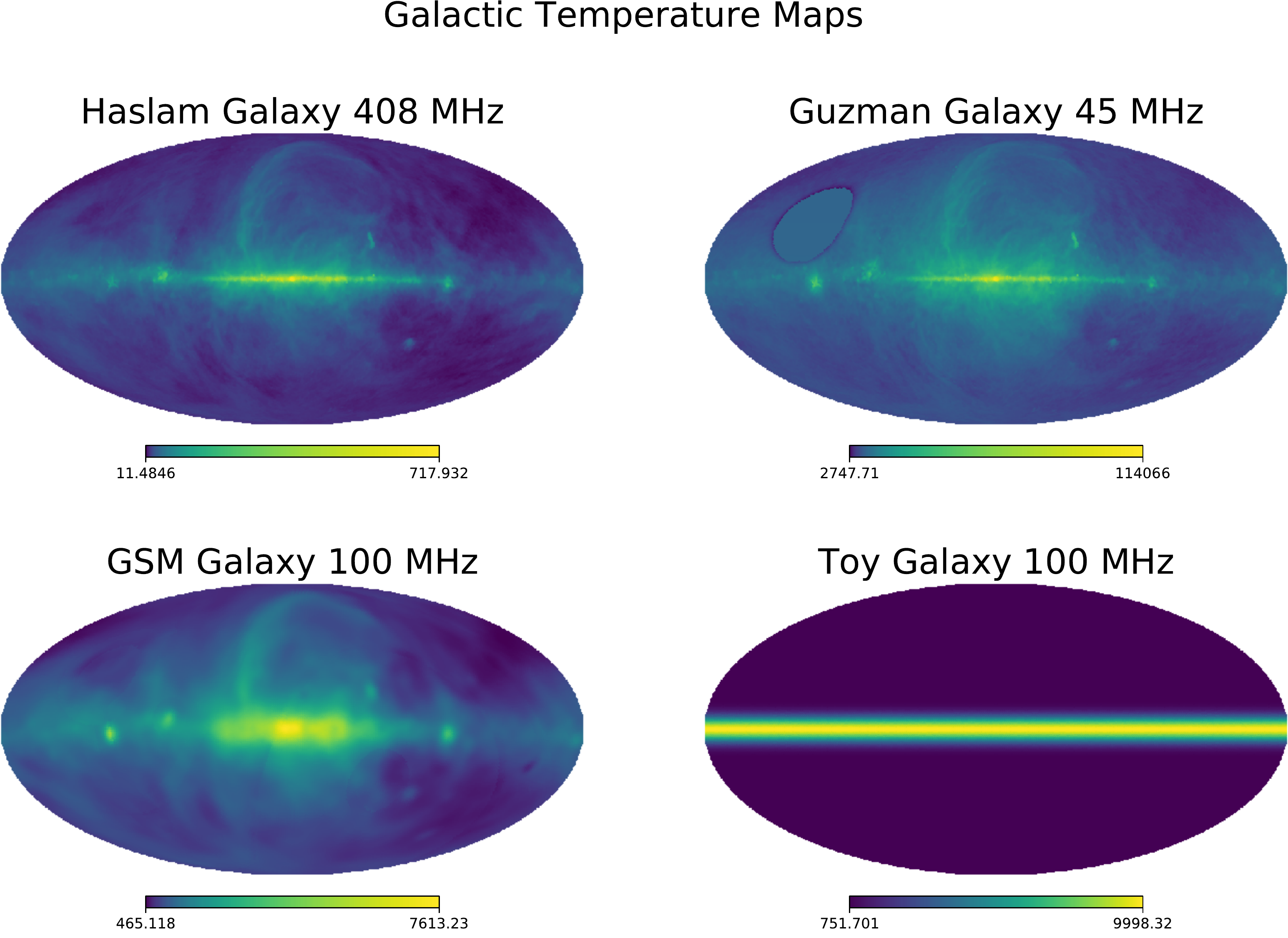}
    \caption{Mollweide projections of sky temperature maps. The Haslam, Guzm\'an, and GSM maps are plotted in logarithmic scale to bring out the fine structures. The hole in the Guzm\'an map has been filled in with the average temperature of the surrounding pixels.}
    \label{fig:tmaps}
\end{figure*}

The other manner in which the foreground may vary, according to the modelling of Equation \ref{powerlaweq}, is in the temperature of the galactic sky, $T_{map}$. As mentioned before, in the frequency range of interest there are two nearly complete sky maps, that of \cite{Haslam:82} and \cite{Guzman:2011}. The GSM model of \cite{Zheng:2017} provides another full-coverage comparison, being an interpolation of the former two with maps at higher frequencies contributing slightly to the interpolation as well. In addition to these, we constructed a temperature map using an overly simplified model of the galaxy which merely captures the most crucial features of the galactic sky: namely, that of one temperature inside the galactic plane, and one outside of it. This map, which we refer to as the Toy Galaxy, consists of a disc in the galactic plane. Mollweide projections of these four temperature maps are shown in Figure \ref{fig:tmaps}.

Summaries of the various spectral index models and sky temperature maps are shown in Table \ref{spectral-index-model-table} and \ref{tab:sky-temp-table}, respectively.

\begin{table*}
    \caption{Spectral Index Models}
    \centering
    \begin{tabular}{ c  c  c }
        \hline\hline
        Spectral Index Model $\beta$ & Symbol & Type \\
        \hline
        Gaussian & G & \\
        Sine-squared & S & \\
        Perturbed & P & \\
        All (Combine Gaussian, Sine-squared, Perturbed) & All &  Analytical \\
        Mozdzen range Gaussian & MG &\\
        Mozdzen range Sine-Squared & MS & \\
        Mozdzen range Perturbed & MP & \\
        Constant & C & \\
        \hline
        Guzman-Haslam (\cite{Guzman:2011}, \cite{Haslam:82}) & GH & \\
        Global-Sky-Model (\cite{Zheng:2017}) & GSM & Interpolated Observational \\
        Long-Wavelength Array (\cite{Dowell:2017}) & LWA & \\
        \hline
    \end{tabular}
    \label{spectral-index-model-table}
\end{table*}

\begin{table}[t!]
    \caption{Galactic Temperature Maps}
    \centering
    \begin{tabular}{ c  c }
    \hline\hline
    Sky Temperature Map $T_{sky}$ & Symbol \\
    \hline
    Haslam Galaxy (\cite{Haslam:82}) & HG \\
    Guzm\'an Galaxy (\cite{Guzman:2011}) & GG \\
    Global-Sky-Model Galaxy (\cite{Zheng:2017}) & GSM \\
    Toy Galaxy & TG \\
    \hline
    \end{tabular}
    \label{tab:sky-temp-table}
\end{table}

\section{Results I:\\ The Monopole-Beam}
\label{Results_1}

In order to understand how the beam distorts the foreground, it is useful to first study the undistorted, or unweighted, foreground---that is, the monopole, where the sky is effectively measured with an achromatic, isotropic beam. This is the ideal case, and constitutes a measurement of the true galactic sky temperature independent of beam effects. Even though no beam is actually involved, we shall refer to this measurement as the ``monopole-beam'' in order to emphasize its relation to the later true beam-weighted foregrounds of Section \ref{Results_2}. The resultant antenna temperature for the monopole-beam is then given by

\begin{equation}
    T_{mon}(\nu) = \frac{1}{N_{pix}} \sum_{i=1}^{N_{pix}} T_{map}(i) \Bigg(\frac{\nu}{\nu_o}\Bigg)^{\beta_{ref}+\epsilon_i}
    \label{monopole_temp},
\end{equation}
where $\beta_{ref}$ is a global sky average reference spectral index perturbed by a unique $\epsilon_i$ at each pixel. The former is generally approximated in the community as $\beta_{ref} = -2.5$. The reference frequency, $\nu_o$, is the frequency at which $T_{map}$ is measured.

This sum of power laws, which we will refer to hereafter as the LinLog (Linear Logarithmic) polynomial for reasons to be explicated in the next section, is the governing equation for the undistorted temperature of the galactic sky, assuming a power law emission component like synchrotron radiation dominates in every pixel. Each of the foreground models we have assembled so far can be substituted into Equation (\ref{monopole_temp}) and summed over the entire galactic sky to give the monopole-beam. 

With 11 different spectral index models (as shown in Table \ref{spectral-index-model-table}) and 4 different sky temperature maps (Table \ref{tab:sky-temp-table}), we can create 44 unique sets of beam-weighted foregrounds, even though in this section the beam is achromatic and isotropic. For each unique spectral index model and sky brightness temperature map combination we calculate thousands of beam-weighted foreground curves using Equation (\ref{monopole_temp}), the frequency range 40 to 120 MHz, and the methods of offset distributions and perturbations delineated in Section \ref{Methods}. The resultant curves for each of the 44 cases are assembled into what we shall refer to hereafter as a training set.

Next, we run SVD on each training set using the pylinex code\footnote{https://bitbucket.org/ktausch/pylinex} (T18) to generate its eigenmodes and singular values. These eigenmodes, as mentioned above, represent the optimal basis to fit every curve in the training set. They are optimal in the sense that they minimize the total RMS residuals produced from fitting every curve in the training set. For example, the first six eigenmodes\footnote{We choose six eigenmodes as a reference level because this is usually sufficient to model the corresponding training set down to a noise level of 1 mK. Ultimately, our pipeline determines the number of modes to use in fitting either through the Deviance Information Criterion (DIC), as discussed in T18, or through the Bayesian evidence (Bassett et al., submitted to ApJ).} (those with the largest singular values) for a monopole-beam training set using all eleven different spectral index models are shown in Figure \ref{fig:monopole_ts_modes} for the Haslam sky temperature map. The eigenmodes are remarkably similar across all spectral index models.

\subsection{A Power Law Digression}

\begin{figure}
    \centering
    \includegraphics[width=0.46\textwidth]{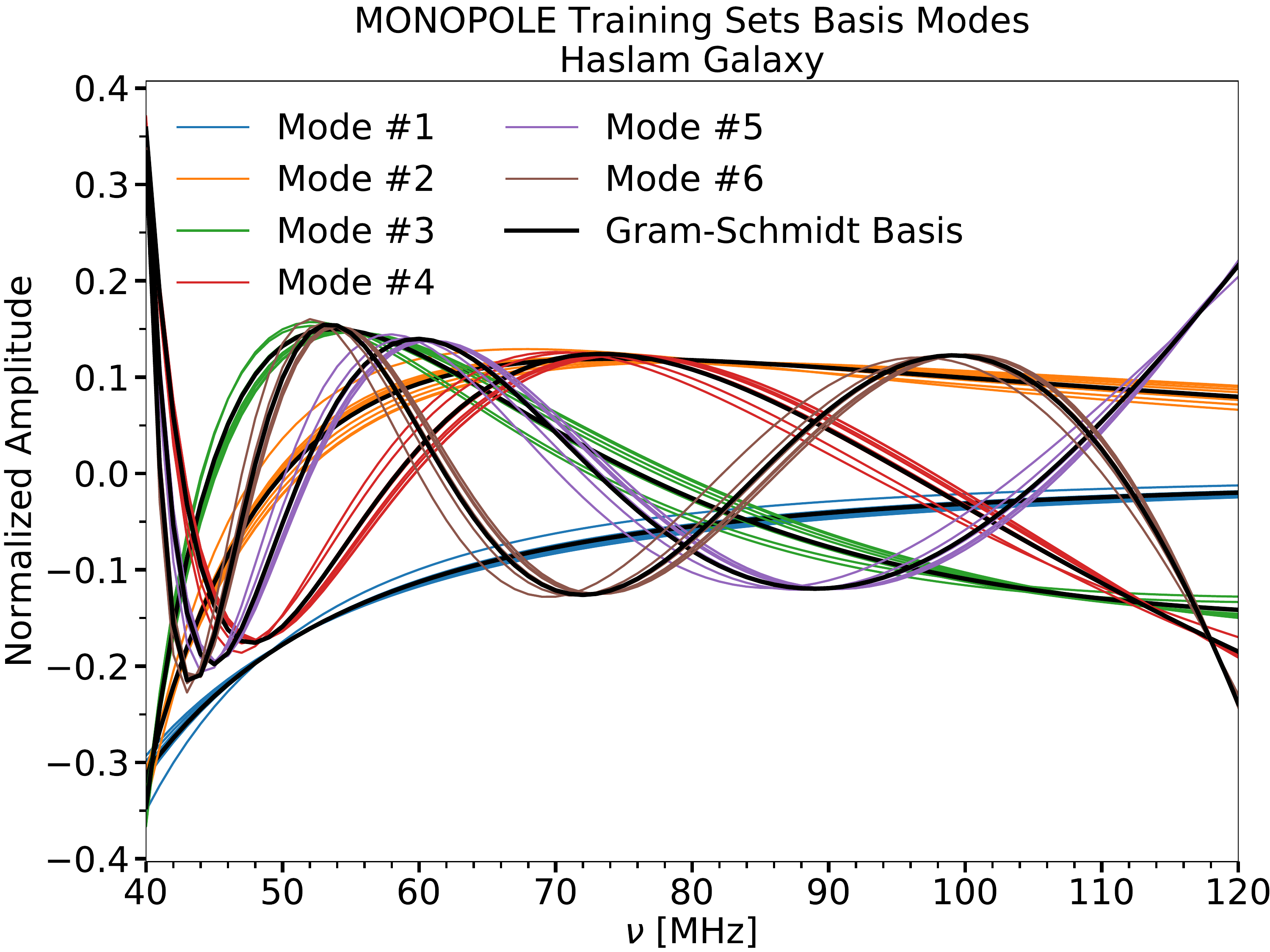}
    \caption{Eigenmodes for a training set using all eleven spectral index models, the Haslam brightness temperature map, and the monopole-beam (achromatic and isotropic). Each spectral index model is plotted together to highlight the similarity between the modes of each model. The black curves show the orthogonalized (via the Gram-Schmidt procedure) form of the LinLog basis given by Equation (\ref{powerlawlogpoly}).}    \label{fig:monopole_ts_modes}
\end{figure}

To shed light on the intriguing yet striking similarities between the training set eigenmodes in Figure \ref{fig:monopole_ts_modes}, it is illustrative to study the Taylor series expansion of Equation (\ref{monopole_temp}), i.e. the monopole-beam, which is in fact Equation (\ref{discreteskytemp}) with $B = 1/N_{pix}$ at all frequencies---an achromatic, isotropic ``beam''. In theory, this is the actual global temperature of the sky that would be measured by a beam without spatial and spectral distortions, the ideal case. Expanding $\epsilon_i$ about zero gives

\begin{equation}
    \begin{split}
        T_{mon}(\nu) = \frac{1}{N_{pix}} \Bigg (\frac{\nu}{\nu_o} \Bigg )^{\beta_{ref}} \sum_{i=1}^{N_{pix}} T_{map}(i)  \Bigg[1 + \ln{ \Bigg( \frac{\nu}{\nu_o} \Bigg)} \epsilon_i \\ + \frac{1}{2} \ln^2{ \Bigg( \frac{\nu}{\nu_o}} \Bigg) \epsilon_i^2 + \frac{1}{6} \ln^3{ \Bigg( \frac{\nu}{\nu_o}} \Bigg) \epsilon_i^3 + O(\epsilon_i^4) \Bigg].
    \end{split}
    \label{powerlawlogpoly}
\end{equation}
Indeed, we recover a common 21-cm galactic foreground subtraction model: that of a power law times a polynomial in logarithmic space (the LinLog polynomial aforementioned). The perturbation $\epsilon_i$ can be both positive and negative, but for a suitably chosen value of $\beta_{ref}$ should certainly be less than unity. Even the largest ranges in spectral index reported at these low frequencies are still limited below this level. See, for example, the ranges stated in Section~\ref{sec:constant-spectral-index-maps}.

\subsection{Gram-Schmidt Basis}
\label{gram-schmidt-basis}

The above digression thus explains why every summation of power laws (within a small spectral index range) can be approximated by a power law multiplied by a polynomial in $\ln{\nu}$ with a spectral index that is essentially the average of the galactic sky's spectral indices, and hence why all of the modes for each spectral index model in the monopole-beam are so similar in Figure \ref{fig:monopole_ts_modes}. Our training sets are merely sums of simple power laws. 

Indeed, the modes produced by SVD are actually just the orthogonalized versions of the terms in Equation (\ref{powerlawlogpoly}). This can be seen if we use the first six terms as ``seed vectors'' for an implementation of the Gram-Schmidt algorithm, which iteratively produces an orthogonal basis from seed vectors by finding and removing the overlap between each seed vector and the orthogonal basis generated from the previous seed vectors, subtracting it, and normalizing \citep[see][]{Greub:1981}. The SVD modes---optimal for fitting the training sets---are, up to a sign, extremely similar to those Gram-Schmidt basis vectors, which are overlaid in black in Figure \ref{fig:monopole_ts_modes}, showing that Equation (\ref{powerlawlogpoly}) is a very good approximation of the monopole-beam foreground when each source has a power law spectrum.

\begin{figure*}
    \centering
    \includegraphics[width=0.46\textwidth]{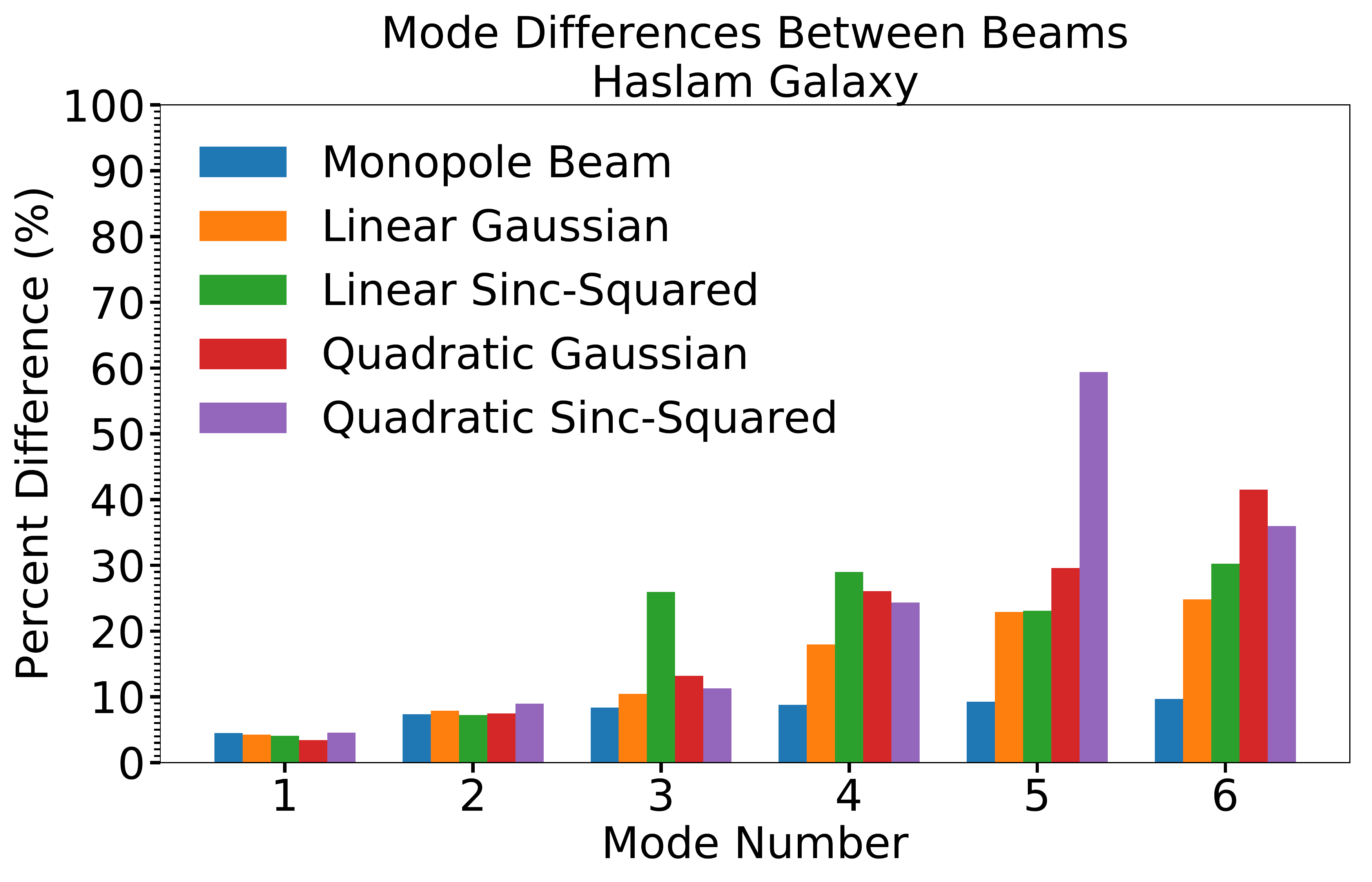}
    \includegraphics[width=0.46\textwidth]{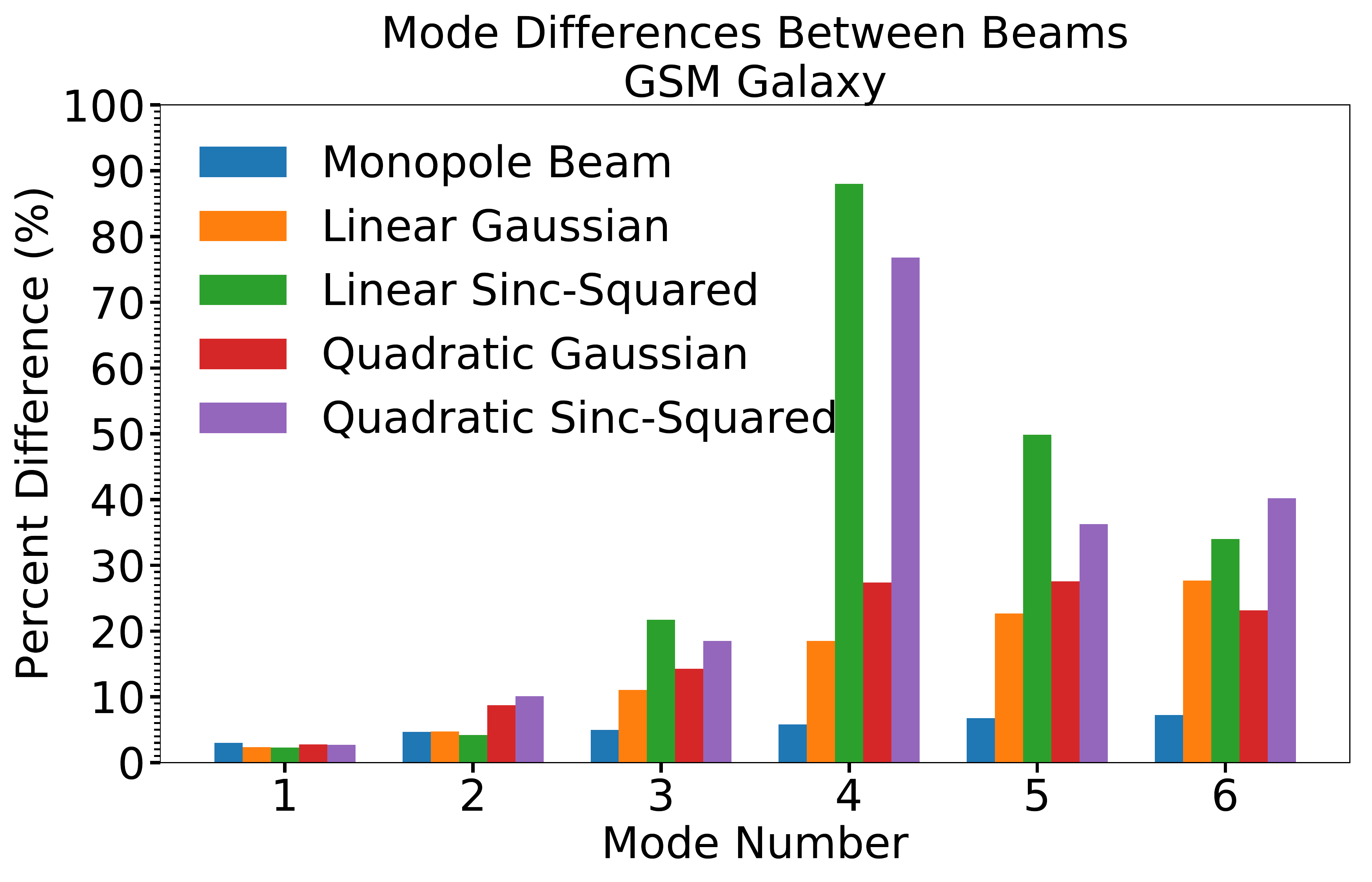}
    \includegraphics[width=0.46\textwidth]{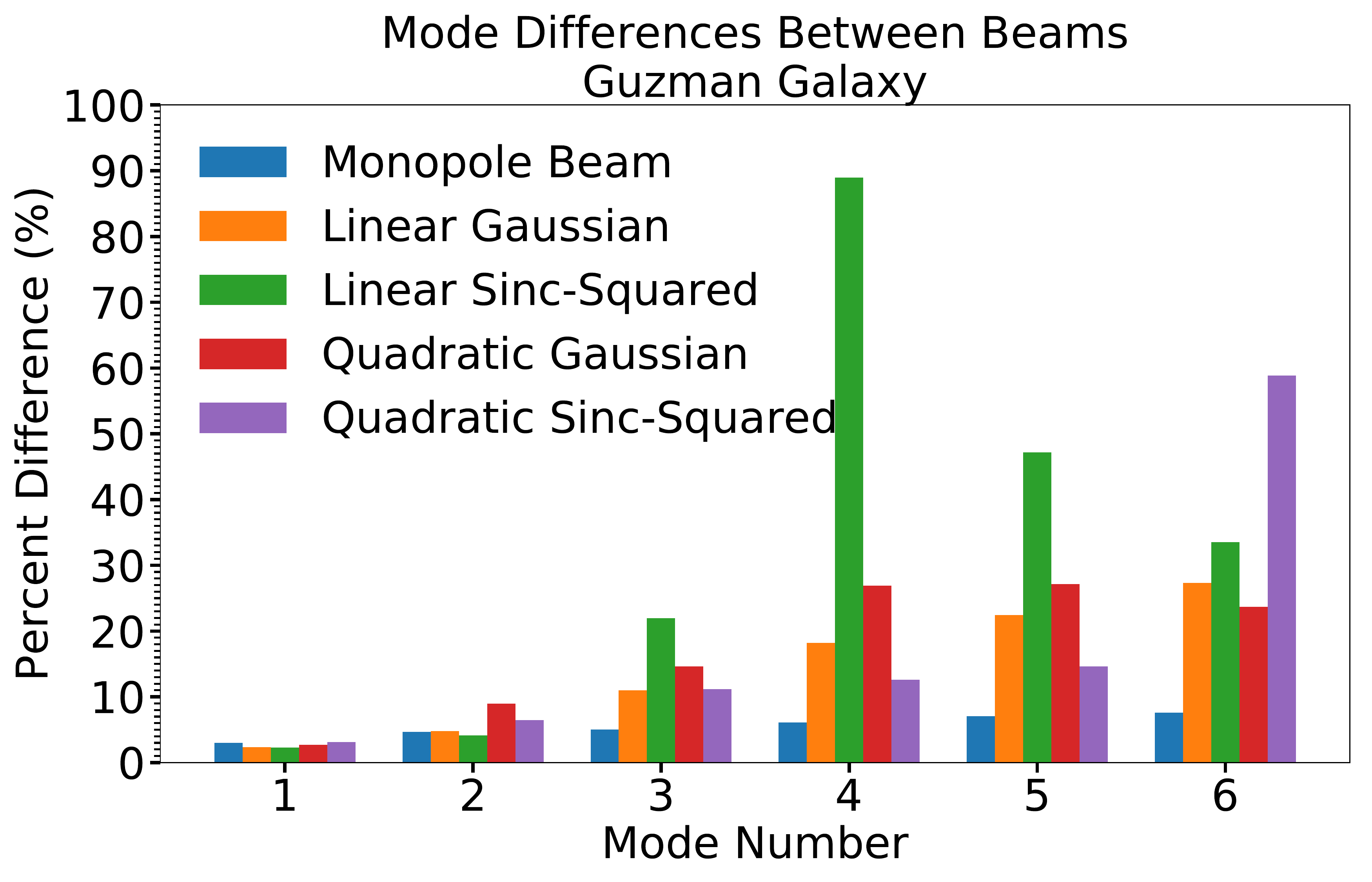}
    \includegraphics[width=0.46\textwidth]{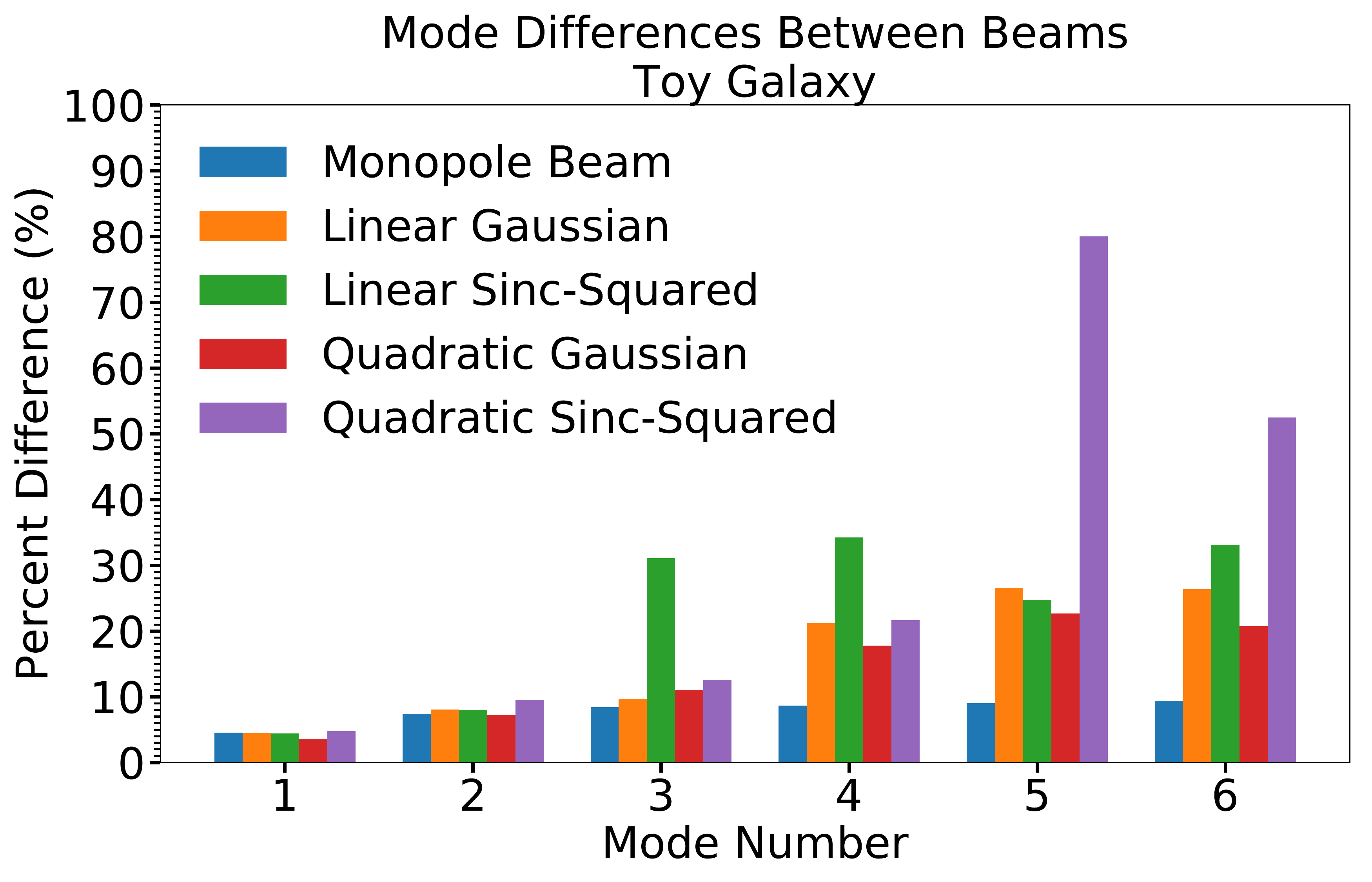}
    \caption{Percent difference in mode amplitude between different galaxy spectral index model training sets for a given beam (including the monopole-beam) and sky brightness temperature map (as shown in each panel's title). The percent difference is calculated by finding the RMS between all 11 spectral index models at a given frequency for a particular mode, and then calculating the RMS across frequencies for that particular mode. This shows that the monopole-beam modes are the least different across galaxy models for each mode, while introducing the subsequent beams changes the optimal modes for each training set according to the galaxy model used to construct it. For the simulations in this figure, the pointing of the beam was the NGP.}
    \label{fig:mode-percent-diff-barplot-hg}
\end{figure*}

\subsection{Optimal Monopole-Beam Eigenmodes}

Overall, the monopole-beam analysis highlights two important results, as follows. First, the optimal modes for modelling the monopole-beam foreground are nearly identical \textit{regardless of the galaxy model used to construct the foreground}, and hence those eigenmodes are not dependent upon the exact spatial and spectral structure of the foreground. In Figure \ref{fig:mode-percent-diff-barplot-hg}, we quantify the similarity in the monopole-beam modes (for each sky brightness temperature map) by calculating the percent difference between spectral index (and thus galaxy) models for each mode across the entire frequency band. This value is found by first calculating the root-mean-square (RMS) between the spectral index models at each frequency for a given mode, and then finding the RMS for that mode across the entire frequency band. For each mode, Figure \ref{fig:mode-percent-diff-barplot-hg} shows that the 11 monopole spectral index models vary between themselves four to six percent on average, thus supporting the qualitative similarities between each model seen in Figure \ref{fig:monopole_ts_modes}.

The second key result of this analysis is that LinLog polynomial models match well the monopole-beam SVD eigenmodes, as demonstrated in Section \ref{gram-schmidt-basis}. Such polynomial models are thus only optimal for achromatic, isotropic beams and are unrealistic models for most experiments.

On the other hand, as we shall demonstrate in the next section, chromatic, anisotropic beams and their weighting upon the foreground, when coupled with the various galaxy models, distort the monopole-beam eigenmodes and consequently the optimal model away from the LinLog polynomial, while becoming significantly dependent on the galaxy model.

\section{Results II: \\ The Beam-Weighted Foreground}
\label{Results_2}

\subsection{Beam Pattern and Chromaticity}

In order to analyze the effect of the beam upon the foreground, we simulate four different beams where the coupling between the spatial structure (i.e. angular shape) and frequency structure (spectral) is varied. The first beam is spatially Gaussian and has a FWHM that varies linearly with frequency:
\begin{equation}
    \text{FWHM}_{\text{Lin}}(\nu) = 115^\circ - 37.5^\circ \left(\frac{\nu}{\nu_r}\right),
    \label{linearfwhm}
\end{equation}
where $\nu_r$ represents a reference frequency, here chosen to be 100 MHz. The first beam's angular structure is given by
\begin{equation}
    B_{\text{Gaussian}} \propto \exp \left\{-\ln{2}\left[\frac{\theta}{\text{FWHM}(\nu)/2}\right]^2 \right\}.
\end{equation}
The second beam is also a Gaussian spatial beam, but with a quadratically varying FWHM so as to examine the effects of its ``spectral curvature'' upon the eigenmodes:
\begin{equation}
    \text{FWHM}_{\text{Quad}}(\nu) = 101.6^\circ - 88.2^\circ \left(\frac{\nu}{\nu_r}\right) + 0.956^\circ \left(\frac{\nu}{\nu_r}\right)^2.
\end{equation}
The Quadratic FWHM chromaticity function was motivated by the EDGES beam, as reported in \cite{Bowman:18}, in the same manner as those utilized in T20a.

\begin{figure}
    \centering
    \includegraphics[width=0.46\textwidth]{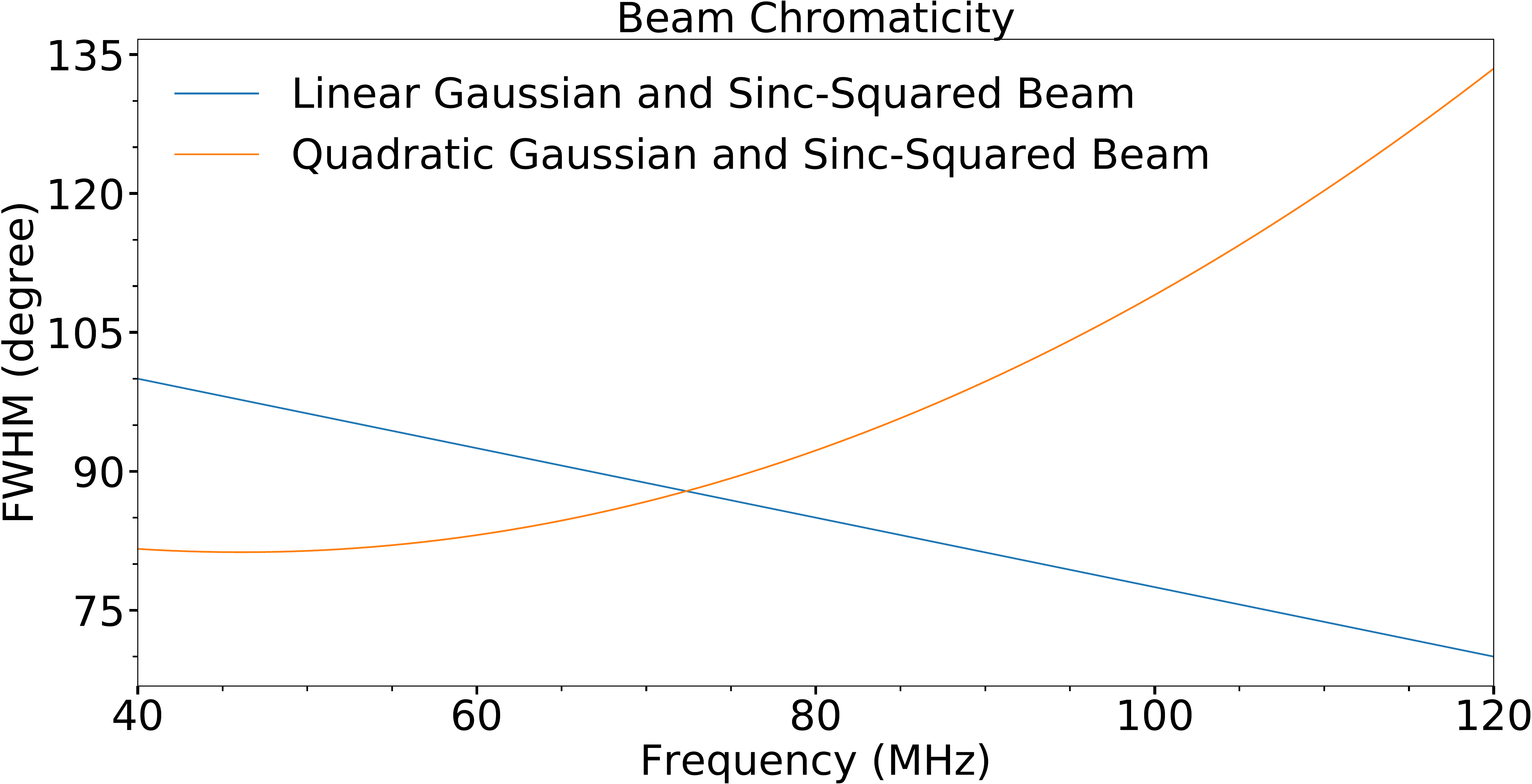}
    \caption{Chromaticity on the FWHM for the Linear Gaussian and Sinc-squared beams (blue line), and the Quadratic Gaussian Sinc-squared beams (orange curve).}
    \label{fig:beam_chrome}
\end{figure}

The last two beams have Sinc-squared beam patterns, with the same chromaticities as the Gaussian beams, one linear and one quadratic, so that comparisons can be made between beams with the same FWHM chromaticity but varying spatial dependence. The angular structure of the Sinc-squared (S) beam is
\begin{equation}
    B_{S} \propto \left[\frac{\sin{(k\theta)}}{k\theta}\right]^2,\ \ \text{ where } \ \ k=\frac{2.783}{FWHM(\nu)}\,.
\end{equation} 
We refer to each of these beam simulations as the Linear Gaussian, Quadratic Gaussian, Linear Sinc-squared, and Quadratic Sinc-squared beam, respectively. The FWHM ranges and frequency dependence for the beam models are shown in Figure \ref{fig:beam_chrome}.

It should be noted that the range of the FWHM is similar for each beam, and that the linear beam has a FWHM which decreases with frequency (blue line in Figure \ref{fig:beam_chrome}), while the quadratic beam (orange curve) increases with frequency. Altogether we have, including the monopole-beam, 5 beam models.

\subsection{Beam Pointing and Location}

\begin{figure}
    \centering
    \includegraphics[width=0.46\textwidth]{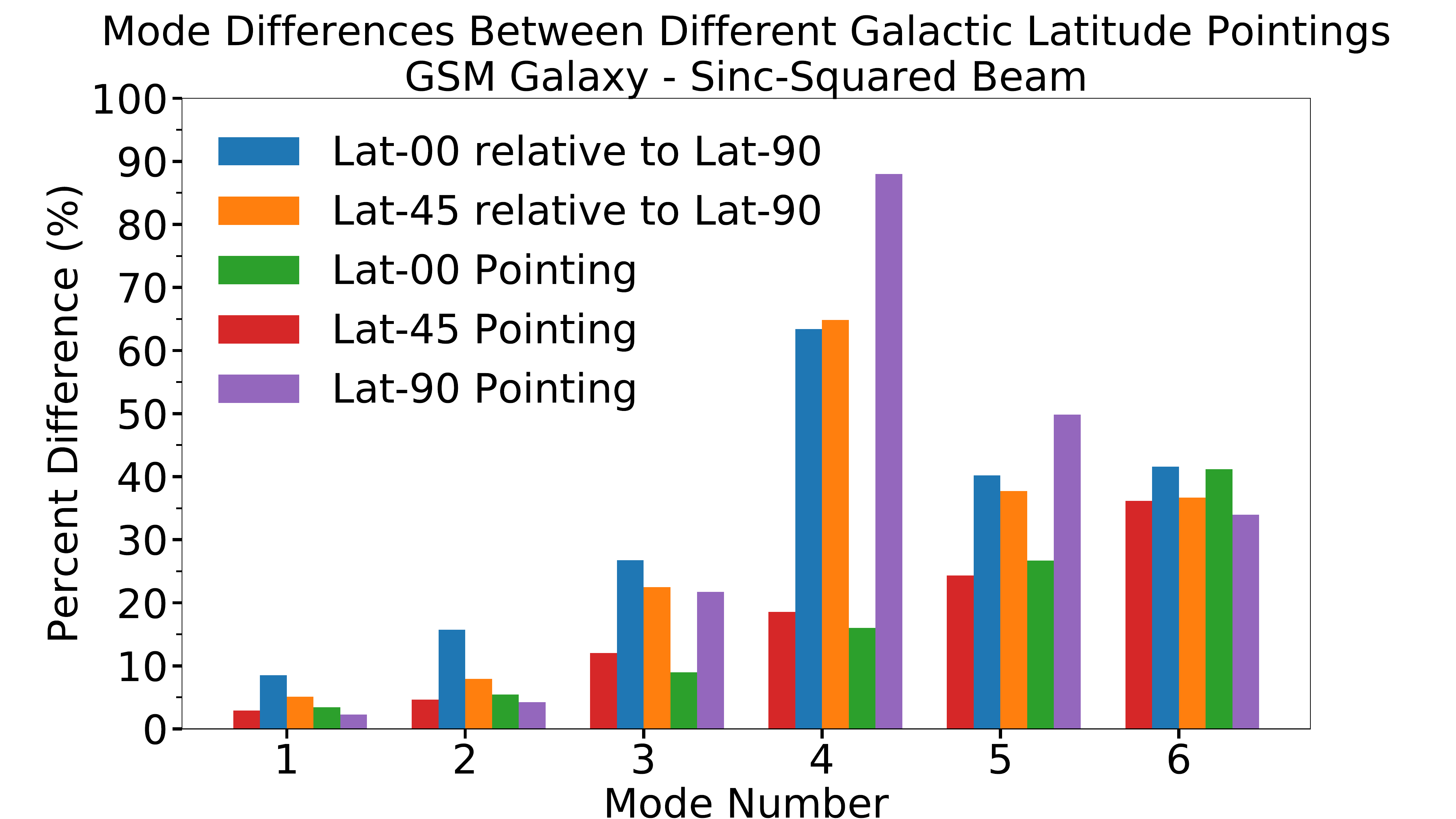}
    \caption{Percent difference between galaxy spectral index model training sets for different Sinc-squared beam pointings (using the GSM sky brightness temperature map): Latitude 90 (purple), the NGP; Latitude 45 (red), mid-galactic latitude; and Latitude 0 (green), the Galactic Center (GC). The blue and orange bars represent the percent difference in the modes calculated for Latitude 0 and Latitude 45 with respect to the Latitude 90 pointing.}
    \label{fig:diffpointings-barplot}
\end{figure}

The beam pointing (for space-based experiments, such as DAPPER) or antenna latitude and zenith-pointing (for ground-based experiments) also change the optimal beam-weighted foreground eigenmodes. Both ultimately depend upon the portion and position of the galactic plane overhead. This can be seen by examining the effects of the pointing (in galactic coordinates) upon the optimal eigenmodes. Figure \ref{fig:diffpointings-barplot} quantifies the differences in a plot similar to Figure \ref{fig:mode-percent-diff-barplot-hg}, showing the mode differences within three separate galactic coordinate pointings: Latitude 90, or the North Galactic Pole (NGP), Latitude 45, and Latitude 0, the Galactic Center (GC). The Latitude 45 and Latitude 0 have modes quite different from the NGP pointing, as can be seen from the orange and blue bars, which are percent differences calculated relative to the Latitude 90 pointing.

\subsection{Beam Corruption of Eigenmodes}
\label{Section-beam-corruption-of-modes}

\begin{figure*}
    \centering
    \includegraphics[width=0.46\textwidth]{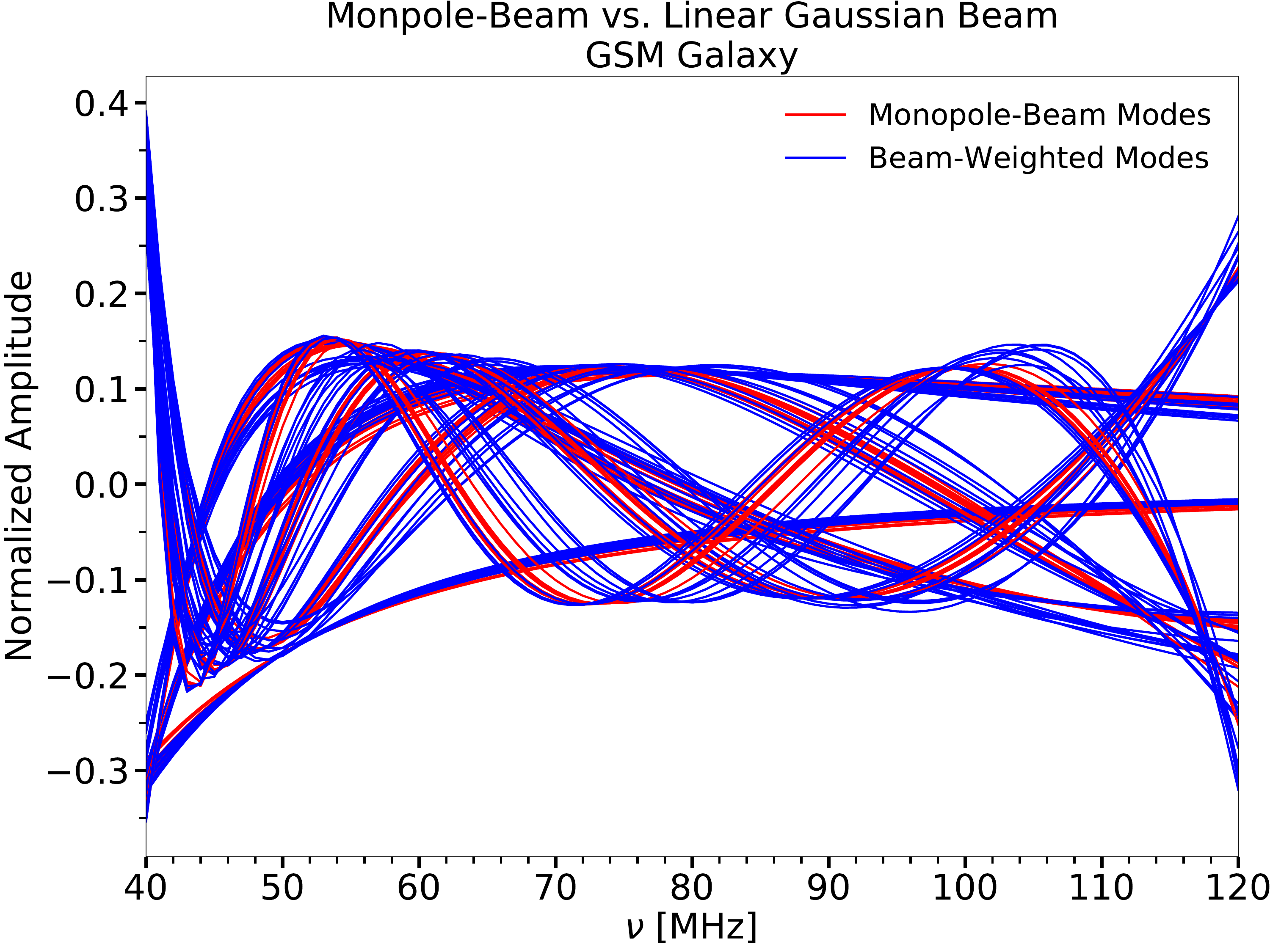}
    \includegraphics[width=0.46\textwidth]{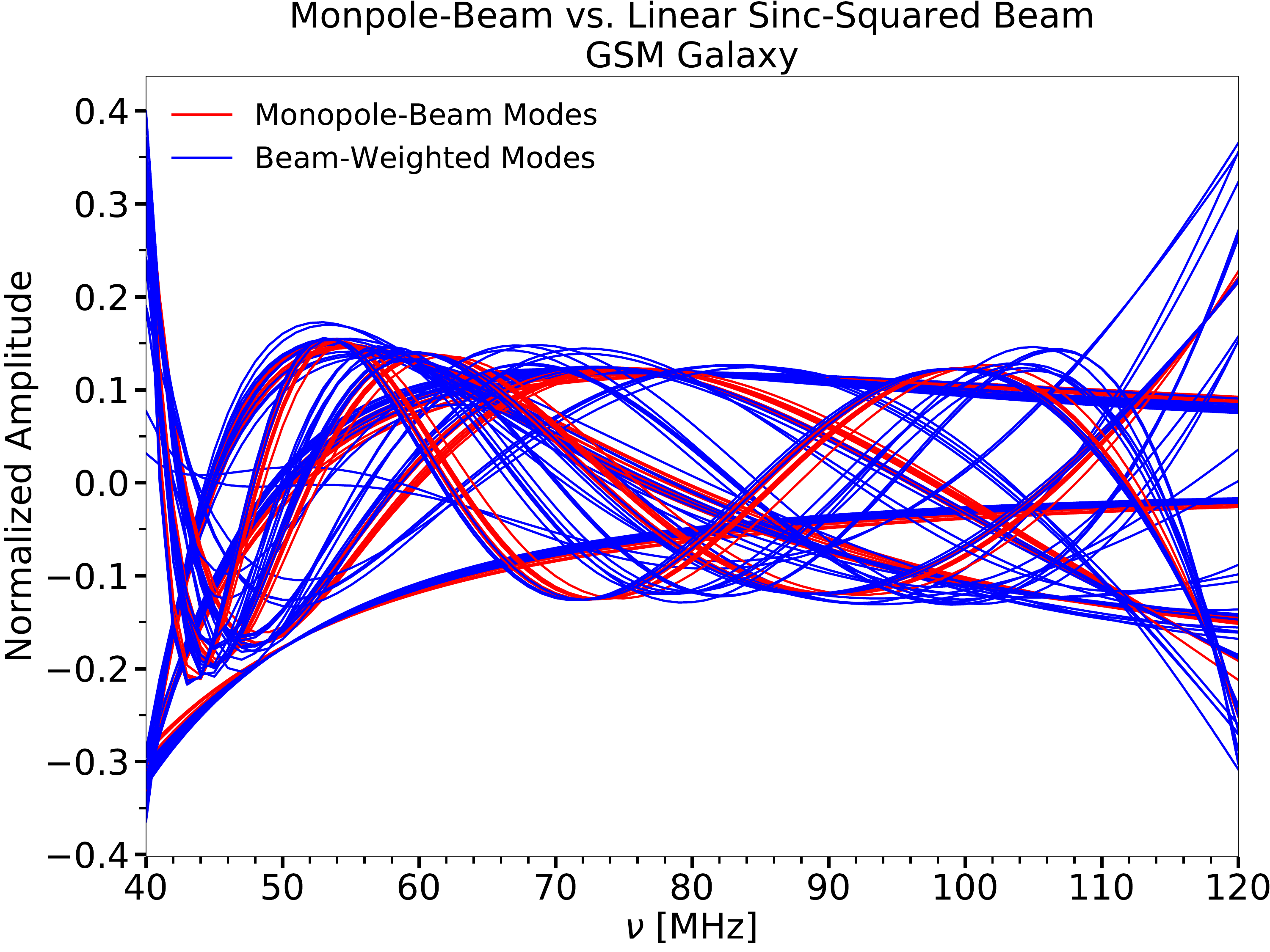}
    \includegraphics[width=0.46\textwidth]{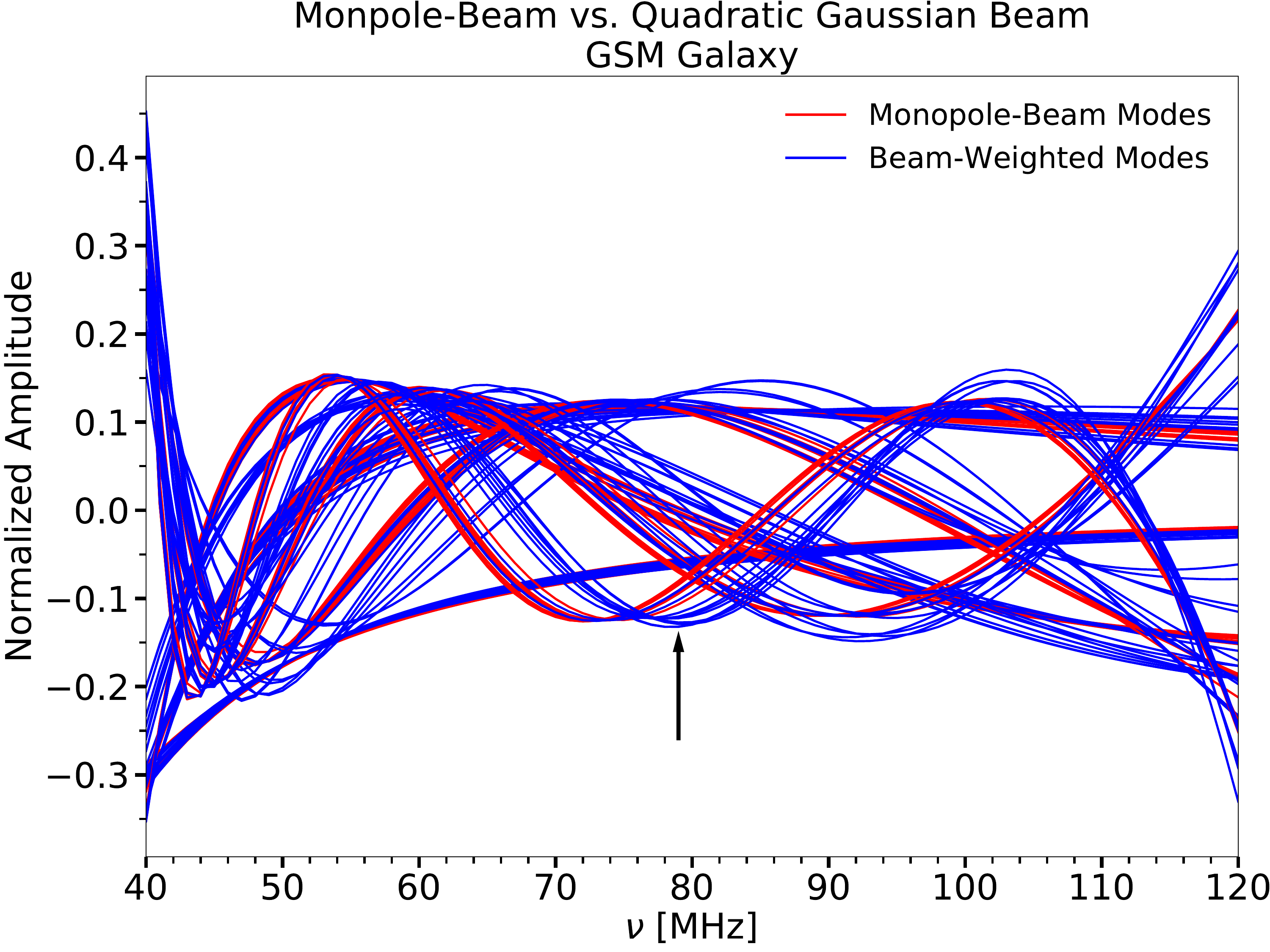}
    \includegraphics[width=0.46\textwidth]{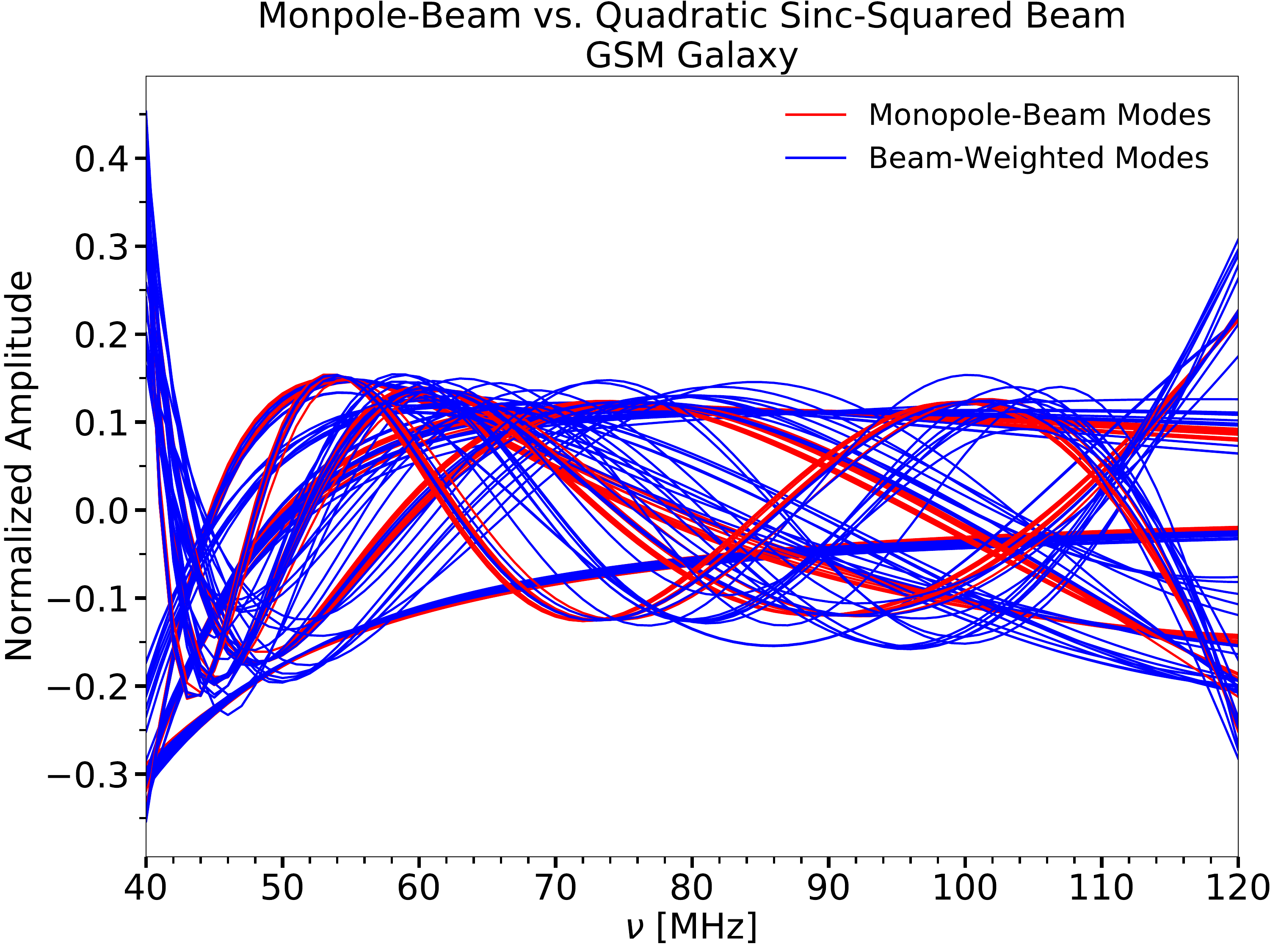}
    \caption{Effects of the beam upon SVD eigenmodes. Blue shows the distorted (beam-weighted) modes, while red shows the monopole-beam modes. The GSM was used as the sky brightness temperature map. The arrow in the bottom left figure denotes the shift from roughly 73 to 79 MHz in the sixth beam-weighted eigenmode exhibited by nearly all of the spectral index models weighted by the Quadratic Gaussian Beam (see Section \ref{Section-beam-corruption-of-modes}).}
    \label{fig:compare_modes}
\end{figure*}

Of all the beam models in Figure \ref{fig:compare_modes}, the Linear Gaussian (top panel) exhibits the least distortion. Figure \ref{fig:mode-percent-diff-barplot-hg} quantifies the differences between spectral index models across frequency for each simulated beam. The Sinc-squared beam's angular-dependence contains side lobes not present in a Gaussian beam, and so parts of the sky normally hidden by the small tail-end weights of the Gaussian beam are brought to prominence (large singular values) with this beam. The resultant eigenmodes therefore depend crucially on the exact spatial and spectral structure of the sky, exposing the differences inherent in each galaxy model. This is why they have split from one another, and are not only shifted spectrally like for the Linear Gaussian Beam.

For a beam with chromatic curvature, like the Quadratic Gaussian or Sinc-squared beam, splitting of eigenmodes between different spectral index models is also displayed. In this case, the beam's FWHM changes rapidly across the band and hence the sources and angular-frequency features of the galactic sky weighted by the beam, resulting in eigenmodes distorted from the monopole-beam foreground. More intriguing still, in connection with the results of T20a using such quadratic beams, the first trough in the sixth beam-weighted eigenmode in nearly every spectral index model is shifted from the monopole value of $\approx$ 73 MHz to $\approx$ 79 MHz (shown by the arrow in the bottom left panel of Figure \ref{fig:compare_modes}). In T20a it was found that such a quadratic beam can produce a false trough at 78 MHz, similar to that reported by EDGES in \cite{Bowman:18}.

We note also the stability of the first few modes across the different beams; in particular, the first mode, which has the greatest singular value and a power law at the reference spectral index, is essentially identical between the monopole and all three beams.

\subsection{Optimal Beam-Weighted Eigenmodes}

The results above show that the beam induces a dependence of the optimal beam-weighted foreground eigenmodes (and thus model) upon the exact spatial and spectral characteristics of the unweighted foreground. These eigenmodes change depending on the proportion of the galaxy overhead, further inducing a dependence on pointing or LST time. A quantitative demonstration of these effects of the beam upon modelling is given in Section \ref{Discussion}. Furthermore, these results reemphasize that the popularly used polynomial models are unsuited to model these beam-weighted foregrounds, as such models are completely agnostic of the non-negligible effects arising from the beams weighting the foregrounds.

Our method of generating beam-weighted foreground modes does not rely upon correcting the effects of the beam by the introduction of an additive or multiplicative factor in Equation (\ref{discreteskytemp}). Instead the effects of the beam are included directly, and the beam and foreground are modelled as a single data component from which we can generate accurate eigenmodes. \textit{We do not need to invoke an absolute model of the beam or the foreground in order to separate them from the 21-cm signal; rather, we need only know all of the possible ways in which the beam-weighted foreground can vary to properly determine the covariances and thus the corresponding uncertainties.} 

Since we found that achromatic beams lead to nearly identical eigenmodes regardless of the foreground model, we infer that it is the interplay of the beam's chromaticity with the foreground which skews the eigenmodes. Chromaticity, then, can be defined as a coupling between the frequency and spatial structure of the beam.

\section{Discussion}
\label{Discussion}

\subsection{RMS Level}

While Figure~\ref{fig:compare_modes} shows qualitative differences between the modes of different beams and spectral index maps, we wish to quantify the differences in the eigenmodes of each training set in a way meaningful to 21-cm signal analyses. For this, we can examine the average value of the residuals produced when the eigenmodes of a particular training set are used to fit every curve from a different training set. That is, given a particular training set, $\boldsymbol{A}$ (for example, a Gaussian spectral index model using the Haslam sky brightness temperature map and a Linear Gaussian beam), we use SVD to calculate its basis eigenmodes, $\Vec{a}$. Then, given a different training set $\boldsymbol{B}$, we fit each of the unique, beam-weighted foreground curves within $\boldsymbol{B}$ using, in this case, 6 eigenmodes from $\Vec{a}$ and calculate the Root-Mean-Squared (RMS) value of the residuals from that particular fit, which we denote as $\text{RMS}_{\boldsymbol{A}\rightarrow \boldsymbol{B}}^{(i)}$, where the index $i$ refers to the $i$-th foreground curve in $\boldsymbol{B}$. After calculating the RMS value for every curve $i$ in $\boldsymbol{B}$, we calculate the total RMS value according to
\begin{equation}
    \text{RMS}_{\boldsymbol{A}\rightarrow \boldsymbol{B}} = \sqrt{\frac{1}{N_{B}} \sum^{N_{B}}_{i=1} (\text{RMS}_{\boldsymbol{A}\rightarrow \boldsymbol{B}}^{(i)})^2}
    \label{average-rms}.
\end{equation}
This total RMS value is what we use to characterize the goodness of fit using the eigenmodes of training set $\boldsymbol{A}$ to model the foreground curves in training set $\boldsymbol{B}$.

\begin{figure*}
    \centering
    \includegraphics[width=0.96\textwidth]{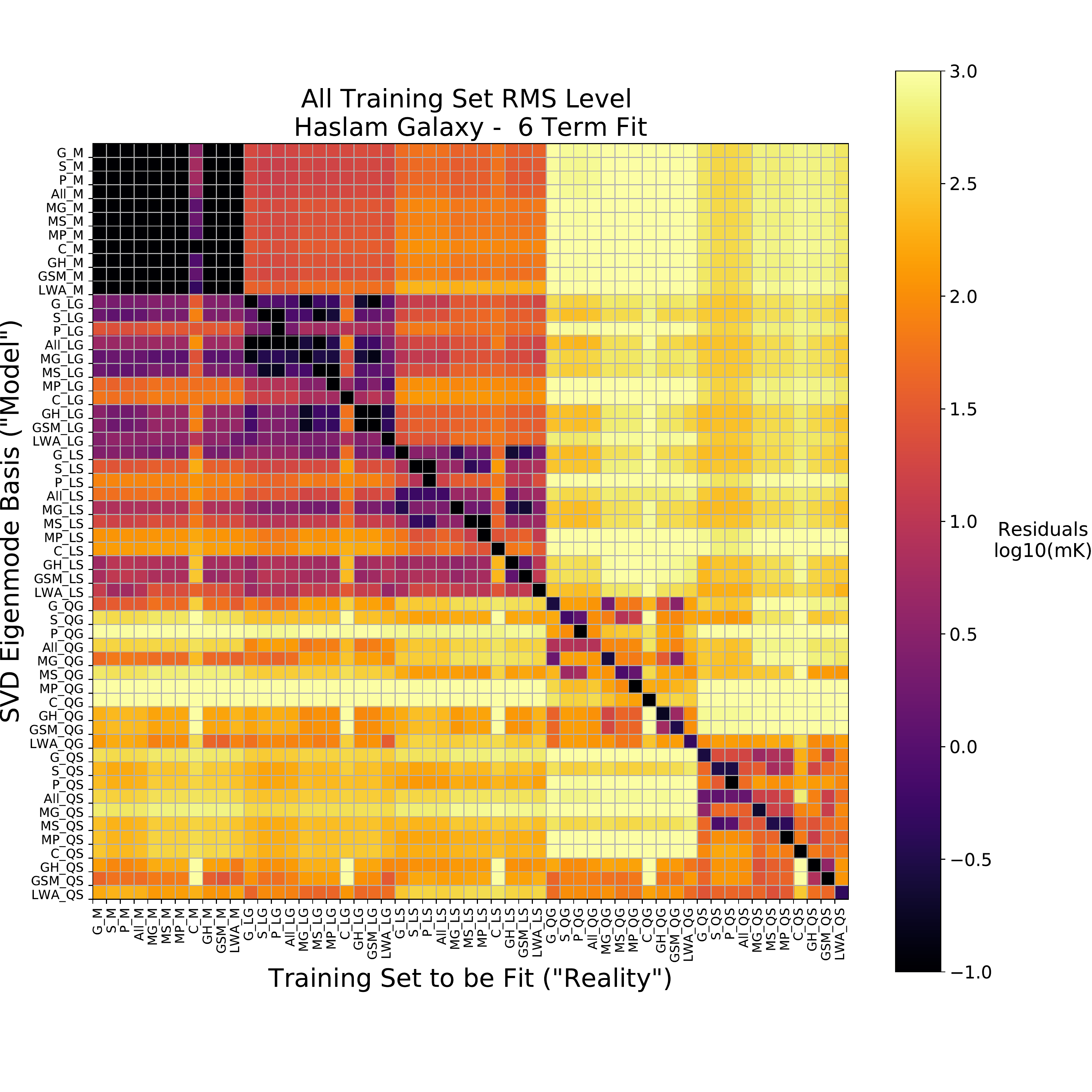}
    \caption{Grid of RMS levels obtained from using the first 6 SVD eigenmodes derived from a given training set to fit every other training set. Each training set is denoted first by its spectral index model followed by the beam used to weight the foreground training set. Spectral Index model labels: G - Gaussian; S - Sine-squared; P - Perturbed; All - Combined Gaussian, Sine-squared, and Perturbed; MG - Mozdzen range Gaussian; MS - Mozdzen range Sine-squared; MP - Mozdzen range Perturbed; C - Constant; GH - Guzm\'an Haslam IO; GSM - Global Sky Map IO; LWA - Long Wavelength Array IO. Beam weighting labels: M - Monopole (achromatic, isotropic beam); LG - Linear Gaussian; LS - Sinc-squared; QG - Quadratic Gaussian; QS - Quadratic Sinc-squared. This grid uses the Haslam Galaxy as the base temperature map.}
    \label{fig:HG_6_term_RMS}
\end{figure*}

The grid of such RMS values in Figure \ref{fig:HG_6_term_RMS} shows what happens when one training set, which we can consider to represent ``reality,'' is modelled using the SVD eigenmodes from another training set, ``the model,'' that differs from that reality in one or more aspects.

\subsection{Differences in Foreground Models}

Much information is contained in Figure \ref{fig:HG_6_term_RMS}, but the more salient facts are these: the beam has the greatest impact upon the beam-weighted foreground eigenmodes. That is, a training set weighted with a particular beam fits all other training sets weighted with different beams to a higher RMS level on average (and thus a greater uncertainty level) than the same beam fits each spectral index model or temperature map contained within its own training set. In most cases, the training sets are too different to allow one training set $\boldsymbol{A}$ to be fit to the same level as a different training set $\boldsymbol{B}$ using the same number of terms.

This is readily apparent from the roughly $11 \times 11$ ``blocks'' of color pervading the RMS grid, where the delineation by beam model dominates the structure (or color). If the beam models were less important than say the spectral index model for determining the average RMS level one training set could fit another down to, then we would expect to see the structure of the RMS grid dominated by striations of lines according to spectral index model. In particular, every other beam-weighted foreground model fits the Quadratic Gaussian and Quadratic Sinc-squared beam-weighted foreground models to the $\sim 10^3$ mK RMS level (yellow color in grids) for 6 terms in Figure \ref{fig:HG_6_term_RMS}.

The spectral index model (and consequently, in the case of the IO models, the sky brightness temperature map) still affects the RMS level produced by the fit as can be seen, for example, in Figure \ref{fig:HG_6_term_RMS} by the orange and yellow horizontal lines corresponding mostly to Constant and Perturbed spectral index models, representing poor fits of comparable RMS level to the Quadratic Gaussian beam-weighted foreground models. Both of these spectral index models consist of uniform shifts in the spectral index itself, resulting in a large range of \textit{reference} spectral index values, $\beta_{ref}$, and thus power laws. The other spectral index models (Gaussian and Sine-squared) consist of curves with perturbations around a single reference spectral index, and it is more difficult to fit many different power laws than it is to fit many small fluctuations around a single power law with a given basis.

As shown in Figure \ref{fig:HG_6_term_RMS}, the SVD eigenmodes of each training set naturally fit their own curves the best, as can be seen by the mostly black diagonals (top left to bottom right). It is worth pointing out that these fits are all achieved down to milli-Kelvin RMS levels or below with 6 terms, which, as mentioned above, is why we have chosen to use 6 terms throughout this paper. Several other training sets are able to fit down the monopole-beam to the noise level of approximately 1 milli-Kelvin (color coded in purple in the RMS grids), in particular the Mozdzen range Gaussian and Sine-squared analytical models, and the Guzm\'an-Haslam and GSM IO maps. The former is due to the fact that the training sets contain smaller spectral index ranges (see Section \ref{sec-analytical-spectral-index-models}) and hence are closer to a power law and the LinLog polynomial.

In the case of the analytical spectral index models, using a training set which combines all of the analytical models into one training set (labelled ``All'' in Figure \ref{fig:HG_6_term_RMS}) produces better fits than the individual training sets comprising it alone do. While using such a combination of training sets will generally increase the uncertainties when using our pipeline (see Papers I-III), due to the increased probability of overlap with the signal model, it will, for the same reason, be more likely to avoid bias (in this context, the difference between the true beam-weighted foreground and the model) if any of the individual training sets encompass the true beam-weighted foreground.

Furthermore, comparing the interpolated observational spectral index models, the LWA spectral index map on average fits all other models worse than those from the GSM and Guzm\'an-Haslam spectral index maps. This is unsurprising, as the LWA not only contains many bright point-sources contributing their own widely varying power laws, which makes it harder to fit with a single basis, but also lacks the large-scale angular frequencies present in the galactic sky (and hence, in the spectral index maps) as it consists of maps created from interferometric measurements.

Indeed, the consequences for LWA model fitting can perhaps be deduced from the form of the principal spectral index map seen in Figure \ref{fig:example_si}, where the higher co-latitudes between 30 and 90 degrees exhibit large spikes in the spectral index. As the low angular frequency components are absent from the LWA maps, the azimuthally averaged spectral index is more sensitive to these point sources and high angular frequency components, flattening the spectral index as compared to the rest of the off-galactic plane sky. This ultimately produces a wider variation in power laws that must be fit down to the noise level with a single basis, which, as we have seen with the Constant and Perturbed models, is more difficult to achieve.

Lastly, our RMS grid results importantly show that the LinLog polynomial represented by the monopole-beam in Figure \ref{fig:HG_6_term_RMS} and often used to model the galactic foreground, produces poor fits for anything other than the monopole-beam, or an achromatic, isotropic beam. Instead, a model specific to the beam being used is necessary to reach the noise level.

\subsection{Summary of Beam-Weighted Foreground Fits and Residuals}

As shown, if one of the features of a beam-weighted foreground's training set is different from reality, the eigenmodes will not represent an optimal basis. Any fitting analysis using sub-optimal eigenmodes will then either contain unaccounted-for beam-weighted foreground systematics, or require a high number of terms to reach the noise level, which will increase the uncertainties in the measurement due to the probability of increased overlap with the signal model. 

\begin{figure*}
    \centering
    \includegraphics[width=0.96\textwidth]{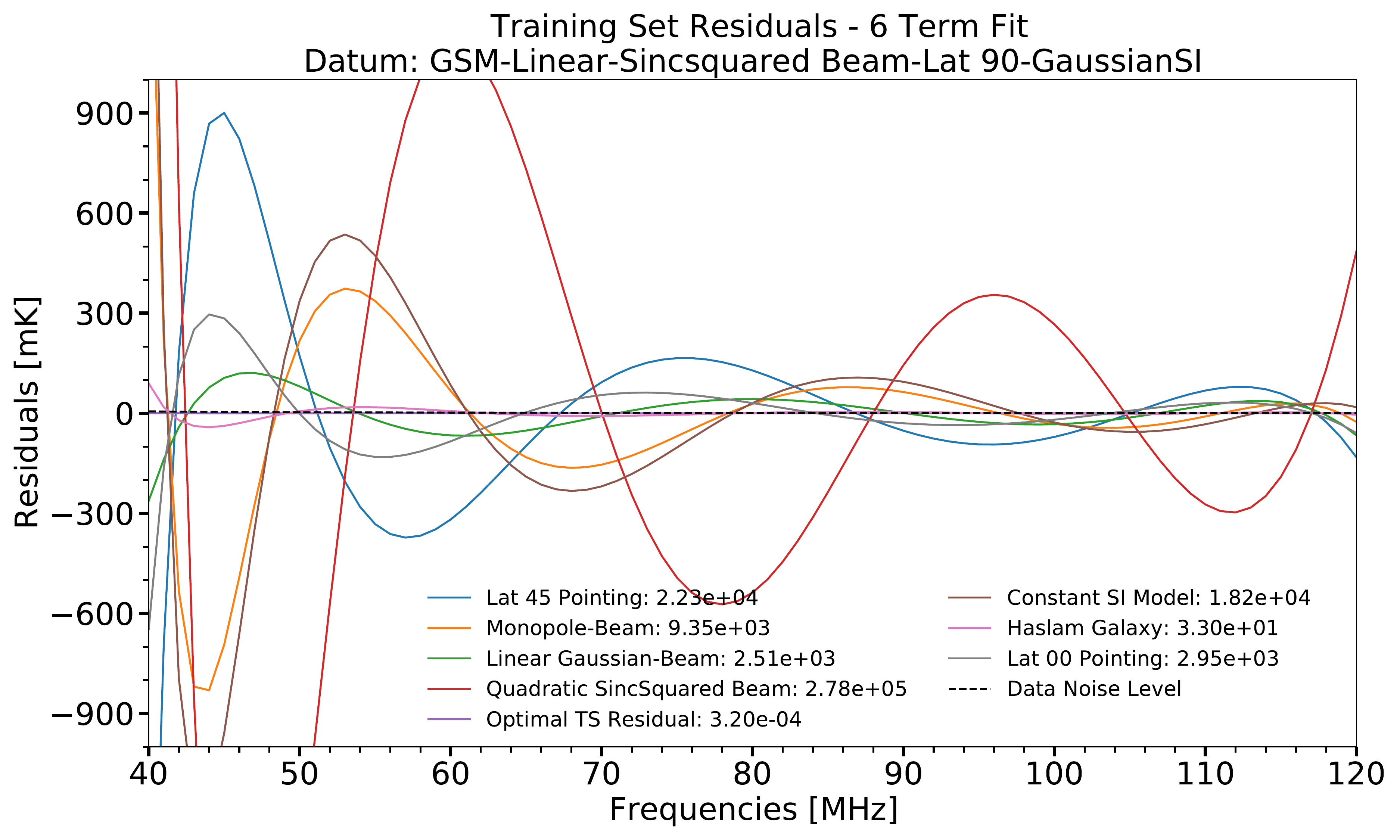}
    \caption{Residuals produced from fitting a datum generated from the GSM sky brightness temperature map, Gaussian spectral index model, and a Linear Sinc-squared beam pointed at the NGP (latitude 90). Each residual curve represents a different six-term fit to the datum using SVD eigenmodes from a training set with a single feature altered from the datum's optimal training set. For instance, the blue curve shows the residual produced when the datum is fit with modes created from a training set with an incorrect pointing (Latitude 45 instead of 90). The dashed line shows the radiometer noise level expected for 800 hours of integration and 1 MHz channel spacing. Only the optimal SVD eigenmodes of the datum's training set are able to fit the datum down to the noise level using six terms. The numbers in the legend next to each curve represent the chi-squared value of each fit, with zero being the expected value (due to the lack of noise in the datum).}
    \label{fig:residual-from-fits}
\end{figure*}

In Figure \ref{fig:residual-from-fits} we show the effects of fitting a datum curve with several models, all of which have a single characteristic tweaked away from the optimal training set SVD basis. In this case, the datum is created using a Sinc-squared beam with Linear chromaticity pointed at the NGP, with no noise added. Each curve is labelled according to which beam-weighted foreground feature has been changed away from the optimal SVD basis and represents the residuals contained after fitting with each non-optimal basis. The optimal basis fit is also shown along with the noise level, the latter of which is generated assuming 800 hours of integration and 1 MHz channel spacing. The number alongside each legend entry represents the reduced chi-squared value of the fit, with the expected value of the fit equal to zero (due to the absence of noise in the datum).

Every non-optimal basis used to fit the datum in Figure \ref{fig:residual-from-fits} contains unaccounted-for systematics, with the Quadratic Sinc-squared beam, differing in chromaticity from the Linear datum, containing the most structure. This figure thus concisely summarizes the need for each beam-weighted foreground feature to be accounted for to produce an optimal model.

\section{Conclusions}
\label{sec:conclusions}

Heretofore, we presented a methodology to understand and generate optimal eigenmodes for modelling the low-frequency beam-weighted foreground by combining analytical models and observational maps with simulations of beams. Using SVD on each unique beam-weighted foreground training set, we can then compare the effects of varying the galactic spectral index, sky brightness temperature map, beam anisotropy/chromaticity, and beam pointing upon the eigenmodes.

We found that when an achromatic, isotropic beam, referred to as the monopole-beam, weights the foreground, the eigenmodes are nearly identical regardless of the galaxy model (spectral index and sky brightness temperature map) used to generate them. Furthermore, we showed that these monopole-beam eigenmodes correspond to the linear polynomial models often used to model the beam-weighted foreground in 21-cm experiments, and that they produce sub-optimal fits and residual systematics when used to model the beams tested here with chromaticity and/or anisotropy. As such, we conclude that polynomial models are only optimal for modelling achromatic, isotropic antennas and are thus generally not realistic models for beam-weighted foregrounds.

When beams with realistic patterns and chromaticites weight the foreground, the chromaticity of the beam couples the spatial dependence of the sky to the spectral behavior of the foreground measurements. Due to this coupling, the optimal eigenmodes for any particular experiment will depend upon the LST times of the measurements, the zenith-pointing of the antenna, and the spatial and spectral characteristics of the foreground model employed, if using a model such as the GSM or interpolating sky brightness temperature maps such as the Haslam or LWA.

Among the models tested, those with the greatest chromatic coupling from the beam had the greatest impact upon the eigenmodes used to model the beam-weighted foreground. When a model with a beam pattern or chromaticity was used to fit another model, or ``reality,'' with a different pattern or chromaticity, the residuals produced with six terms were on the order of $\sim 10-10^3$ mK. In contrast, when a model with a particular beam simulation was used to fit itself, residuals on the order of 1 milli-Kelvin were produced with six terms in the majority of cases tested.

A key lesson from this study regarding the design of 21-cm cosmology experiments is that, among the variations studied here, the relative knowledge of the antenna beam (pattern, chromaticity, pointing) is the most impactful aspect of a  proper characterization of a beam-weighted foreground model's eigenmodes using a training set on its own. It is important to note, however, that for global 21-cm signal extraction using training sets for the foreground and signal simultaneously, the overlap between these training sets plays a central role (see {Section}~\ref{sec:introduction}). The accuracy level required for each feature in the training sets to properly extract the signal may vary depending on the level of this overlap. For example, if the foreground and signal were orthogonal, 10 milli-Kelvin inaccuracies in the foreground model may not significantly impact results. However, when the overlap is large, these same inaccuracies could cause significant biases in the extracted signal.

Using training sets allows one to directly incorporate the effects of the beam into the eigenmodes describing the beam-weighted foreground model. This methodology can be applied to any experiment with a given beam structure, antenna location and pointing. Once a reference level for the beam, such as a simulation of the pattern and chromaticity, is chosen, the methods of this paper can be employed to understand what levels of variation in the training set around the reference level are acceptable when constructing an accurate beam-weighted foreground model. Even for experiments with highly chromatic beams and/or large side lobes, optimal eigenmodes to fit the beam-weighted foreground down to the noise level without beam systematics can be generated as long as the training set accurately represents the sky/experiment. While this work focused on unweighted foreground fluctuations, a complete training set would include both foreground and beam fluctuations (a study of which, is left for future work). To determine whether a training set encompasses the beam-weighted foreground measured by an actual experiment, goodness-of-fit tests have been devised (Bassett et al., submitted) to detect failure in properly modelling a given data component, and thus determine when the corresponding training set must be improved for a more accurate representation.

The results and methodology of this paper can be utilized to produce accurate beam-weighted foreground training sets for any well characterized global 21-cm experiment. Future work will do so for ongoing and planned experiments such as CTP and DAPPER.

\acknowledgements{We would like to thank the referee for insightful and helpful remarks. This work was directly supported by the NASA (National Aeronautics and Space Administration) Solar System Exploration Virtual Institute cooperative agreement 80ARC017M0006. This work was also supported by NASA under award number NNA16BD14C for NASA Academic Mission Services.}

\bibliographystyle{yahapj}
\bibliography{references}

\end{document}